\documentclass[%
article
preprint,
notitlepage,
superscriptaddress,
showpacs,
preprintnumbers,
bibnotes,
amsmath,
amssymb,
aps,
floatfix,
]{revtex4-1}

\usepackage{setspace}
\usepackage{natbib}
\usepackage{graphicx}
\usepackage{dcolumn}
\usepackage{bm}
\usepackage[utf8]{inputenc}
\usepackage[T1]{fontenc}
\usepackage[english]{babel}
\usepackage{amsmath}
\usepackage{amsfonts}
\usepackage{amssymb}
\usepackage{commath}
\usepackage{makeidx}
\usepackage[caption=false]{subfig}
\usepackage{soul}

\DeclareMathOperator{\csch}{csch}

\DeclareMathOperator{\Tr}{Tr}

\DeclareGraphicsExtensions{.pdf,.png}
\graphicspath{{./figs_pre/}{./}}
\usepackage[usenames,dvipsnames]{color}   

\usepackage[normalem]{ulem} 

\def\mean#1{\left< #1 \right>}

\begin{document}

\title{A plausible mechanism  of muscle stabilization in stall conditions}

\author{Hudson Borja da Rocha}
\email{hudson.borja-da-rocha@polytechnique.edu}
\affiliation{
CIRB, CNRS UMR 7241, INSERM U1050, Collège de France and PSL Research University, F-75005 Paris, France }
\author{Lev Truskinovsky}%
\email{lev.truskinovsky@espci.fr}
 \affiliation{
 PMMH, CNRS--UMR 7636 PSL-ESPCI, 10 Rue Vauquelin, 75005 Paris, France}

\date{\today}
\begin{abstract}
We address the well-known limitation of the  Huxley and Simmons 1971 (HS) model. It is a statement that at physiological value of stiffness in the actomyosin complex,  the distribution of the myosin motors becomes microscopically uniform (all the motors are either in pre- or post-power stroke conformation) after an infinitesimal displacement from the stall (isometric contractions) conditions. Such uniform behavior at the fiber level would generate a negative slope in the $T_2-\delta$ relationship (in the nomenclature of the HS paper), not observed experimentally. This negative slope means inhomogeneity of the macroscopic sarcomere configuration, which is also not observed. To address this controversial prediction of the HS theory, we explore the possibility that the slope of the  $T_2-\delta$ curve is, in fact, positive due to an  interaction between neighboring cross-bridges. 
We show that such interaction can potentially destabilize the uniform configurations (all pre or all post) by making the non-uniform configurations energetically preferable.  We argue that, despite the presence of other factors, which can in principle also ensure the microscopic inhomogeneity of cross-bridge configurations, the implied interaction is an important player in muscle mechanics. 
\end{abstract}
\maketitle

\section{Introduction}
\label{intro}

An isometrically constrained skeletal muscle reaches the \emph{stall conditions} when it is fully tetanized. This physiological regime is known as the state of \emph{isometric contractions}. The  mechanical behavior in this state is usually studied by analyzing the response of a tetanized muscle  to  \emph{abrupt} mechanical perturbations   \cite{CT_ROPP_2018,HILL1974267, HS71,Podolsky1960}. From such tests, we now have  a detailed picture of the power stroke machinery responsible for the so called fast force recovery:  a transient passive response that does not involve the detachment of myosin heads and the attendant ATP consumption \cite{Irving02}.

\begin{figure}[ht]
\centering
\includegraphics[scale = 0.9]{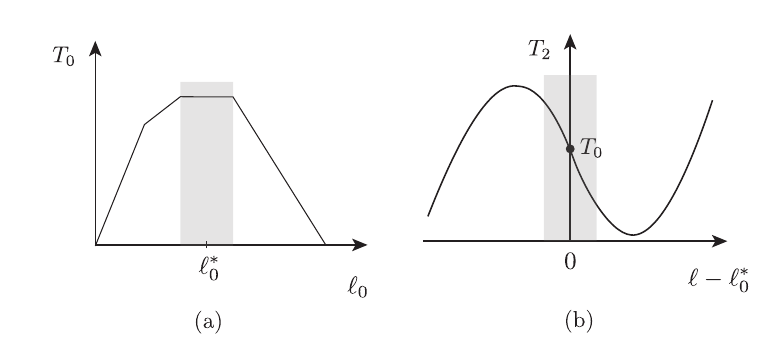}
\caption{(a) Isometric tetanus: actively generated force $T_0$ as a function of the reference length $\ell_0$. (b) Quasi-equilibrium states $T_2$ reached from the stall state $T_0$  by  fast shortenings or stretches. Highlighted are  the main physiological regimes. }
\label{fig:Isometric}
\end{figure}

\begin{figure}[ht!]
\centering
\includegraphics[scale = 1]{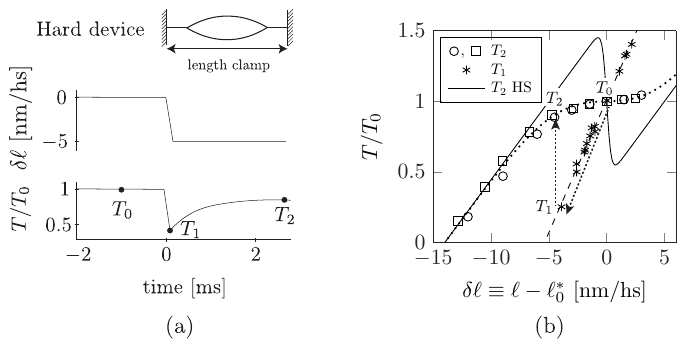}
\caption{  (a) Fast transients in mechanical experiments on single muscle fibers. In response to  the length step $\delta \ell$,  the tension first drops from $T_0$ to $T_1$ and then   recovers  a fraction of the initial tension, reaching the level $T_2$.  (b)  The recovered tension   $T_2$ as a function of the imposed stretch  $\delta \ell$.  Solid curve shows the predictions of the   HS model with realistic  parameters:  $a = 10$nm, $\kappa_0 = 2$pN/nm, $T = 277.15$K, $v_0 = 24.4$pN \cite{CT_ROPP_2018, borjadarocha_thesis}. Data extracted from \cite{Huxley2000, Ford1981, Brunello_PNAS_2007}}
\label{fig:FastRec}
\end{figure}

In the associated experiments,  a muscle is held at the extremities (length clamp or hard device loading)  in an appropriate physiological solution while being electro-stimulated. Under these conditions, it actively generates a  stall force (isometric tension)  $T_0$, which depends on sarcomere length $\ell_0$,  see the schematic picture in Fig.~\ref{fig:Isometric}(a).  The muscle is then shortened (or stretched) by a fixed amount, and the generated tension is measured. For instance, if a (negative) displacement  $\delta \ell=\ell-\ell_0^*$  is applied to a  muscle, isometrically contracting at length  $\ell_0^*$,  the tension first drops to the level $T_1<T_0$,  but then recovers in a few millisecond timescale to the higher level $T_2 $, see the schematic picture in Fig.~\ref{fig:Isometric}(b).  A more detailed picture of the actual experiment together with the available experimental measurements is shown in Fig. \ref{fig:FastRec} (a,b).

While the initial response $T_1$ can be clearly attributed to  elasticity of myosin heads (cross-bridges) \cite{HS71, Lombardikappa}, the recovered tension $T_2$  was shown to result from a conformational transition inside actin-bound myosin heads. During this re-equilibration transition, some of the cross-bridges switch (or fold)  from pre- to post-power stroke state \cite{Irving92, Irving02}. The resulting massive conformational change  is believed to be a purely mechanical process as it does not involve the un-binding of myosin heads from actin filaments\cite{Sheshka2020}.  

Since the homogeneous state of isometric contractions is a  stable equilibrium for the whole muscle fiber, the corresponding global elastic stiffness $dT_2/d\ell$ should be positive, and this is indeed what was observed in the pioneering experiments of Huxley and Simmons (HS) \cite{HS71}, see Fig.~\ref{fig:FastRec}(b). They proposed a theoretical explanation for their measurements, arguing that the elementary force-generating units contain at least two structural elements connected in series \cite{HS71}. The first one is an elastic spring responsible for generating the tension $T_1$, which can be modeled as Hookean. The second is a bi-stable power stroke element that can be either in pre- or in post-power stroke position. HS modeled such element as a hard-spin variable representing the angular position of the myosin head with respect to actin. 
 To ensure a collective response,  individual myosin heads were assumed to be interacting non-locally, through rigid actin and myosin filaments.  
  The resulting model   placed   the thermomechanical behavior of  force generating units (half-sarcomeres) in a chemo-mechanical framework  with  pre- and post-power stroke conformations presented as chemical states.   To obtain analytical results, HS used the Kramers approximation, but their model was recently generalized to account for the full-scale Langevin dynamics \cite{MT10,  Marcucci2010,  CT_PRE, CT_JMPS}.

The hard spin HS model with  effectively paramagnetic-type mechanical interactions could reproduce the experimentally measured tension $T_2$ only if the stiffness of myosin heads was underestimated.   If the correct value (2.7$\pm$0.9 pN.nm$^{-1}$, \cite{Brunello2014, Lombardikappa, Linari19982459})  is used, the HS model leads to the controversial prediction that the free energy of a half-sarcomere is a \emph{non-convex} function of strain, which means, in particular,  that the effective stiffness in the physiological stall force regime is \emph{ negative } \cite{HT96,Vilfan_BioPhys_2003, CAT_PRL}, see Fig.~\ref{fig:Isometric}(b) and Fig.~\ref{fig:FastRec}(b). Behind such non-convexity is the insistence of the model on highly coherent mechanical response of interacting cross-bridges, which essentially precludes  mixing of pre- and post-power stroke conformations. In other words,  given the value of the environmental temperature, the assumed  mean-field elastic interaction unambiguously  points toward necessarily \emph{coherent} response in the state of isometric contractions (stall conditions) with all cross-bridges  either in the pre-power stroke or all in the post-power stroke conformation. 

The  non-positive-definiteness of the tangential elasticity in the constitutive model  can potentially trigger spatial inhomogeneity at the scale of the whole muscle fiber.   In particular,  spatial inhomogeneity should be present in muscle fibers modeled as series connections of half-sarcomeres \cite{Vilfan_BioPhys_2003,Puglisi2000}. Individual half-sarcomeres will be then either entirely in pre- or post-power stroke configurations, which can be interpreted as a coexistence of pure phases. Even though this would lead to a  realistically flat $T_2$ curve in the regime of isometric contractions, the    homogeneous (affine) configurations will be unstable which suggests that the  muscle  fiber will  be structurally highly inhomogeneous \cite{Vilfan_BioPhys_2003, Puglisi2000, Rassier2010}. The fact, that such inhomogeneity of 'muscle material' has not been observed in experiments, stimulated theoretical efforts to account for physical factors ensuring the mixing of the  pre- and post-power stroke configurations at the scale of individual half-sarcomeres.  

Different mechanisms facilitating such mixing have been proposed in the literature. Efforts to \emph{suppress} the instability of an affine response have started with the 1996 paper of Huxley and Tideswell (HT) \cite{HT96}, who suggested that the inter-mixing can be explained by the inherent randomness encapsulated in the actomyosin machinery. More specifically, the authors introduced the idea of a random attachment of the myosin motors along the actin monomer, resulting in a statistical distribution of the attachment positions spread over almost half of the actin diameter (5.5 nm).  This assumption is realistic as the experiments involving electron microscopy \cite{Reedy_Taylor_PlosOne_2012, Tregear20043009}  and X-ray diffraction \cite{Irving02, Reconditi2006} reveal a much more complex geometrical structure of the actomyosin complex than in the original HS model. Thus, myosin binds to actin only in specific target zones, which are not commensurate with the spacing of myosin heads. As a result, different binding sites are reached at different angles \cite{Tregear98, Tregear20043009}.  As it was first realized by HT\cite{HT96}, the implied quenched disorder will compromise the coherent response in stall conditions.  More recently, several other models addressed the same experimental data and successfully reproduced the positive slope of the  $T_2$ curve, even though often without placing emphasis on the quenched disorder incorporated to overcome the original limitation of the HS model, see  \cite{Piazzesi1995, Smith2008, SmithMijailovich2008, Caremani2015} and the references therein. The implications of the presence of disorder in the myosin attachment positions have been recently studied systematically in \cite{BRT_PRL_2019}.  It was shown that having a structural disorder in the positions of myosin molecules relative to attachment sites can effectively \emph{convexify } the elastic response and therefore flatten the $ T_2$ curve.  The rather unexpected result obtained in \cite{BRT_PRL_2019} was that a realistic disorder places a typical muscle system close to a thermodynamic critical point \cite{CAT_PRL}. Despite these fascinating predictions, the idea of the functionality of geometrical disorder in the otherwise highly regular, crystal-type muscle architecture remains debatable. For instance, while the HT model postulates a Gaussian disorder with a \emph{particular} variance, the realistic value of such temperature-like parameter is obscure; moreover, the actual statistical nature of the implied geometrical frustration is unknown. It is then possible, but not certain, that the existing incommensuration is sufficient to ensure the positive slope of the $T_2$ curve. The situation is somewhat similar to the original question faced by HS,  of whether the value of the ambient temperature is sufficient to convexify their free energy.

Local inter-mixing of the pre- and post-power stroke conformations inside a single half-sarcomere can also be facilitated by an \emph{active} process. As shown in \cite{SRT_PRE_2016}, the presence of \emph{correlated} external excitations can generate positive active stiffness in the system, which otherwise has   negative passive stiffness.  In particular,  such  ATP-induced  stiffening  can overcome the negative stiffness  of  the original HS model, making the affine response of muscle fibers stable. Under this hypothesis, the coherency of the mean-field-dominated configurations can be  broken because each element would be forced by active driving to stay away from a single energy well.  Note that the implied active mixing of cross-bridge conformations is different from the passive 'melting' taking over at high temperatures. Active driving, however,  is energetically costly as it requires   incessant ATP consumption. Therefore, it is still difficult to say how realistic is such stabilization scenario.

Given this uncertainty, it is natural to explore other possibilities for passive (rather than active)  stabilization of affine response of muscle fibers in stall conditions. With this aim in view, we put forward in this paper a \emph{plausible hypothesis} that the undesirable, highly coherent mechanical response of muscle half-sarcomeres may be compromised by the destabilizing short-range interaction between individual myosin heads.  Such interaction would then compete with the stabilizing long-range elastic effect of the apparently semirigid filaments. To support this hypothesis, we recall in Sect. 2 some experimental observations and discuss the  \emph{physical mechanisms}  which may ensure the destabilizing character of the short-range interaction between neighboring myosin heads.

\begin{figure}[ht!]
\centering
\includegraphics[scale = 1]{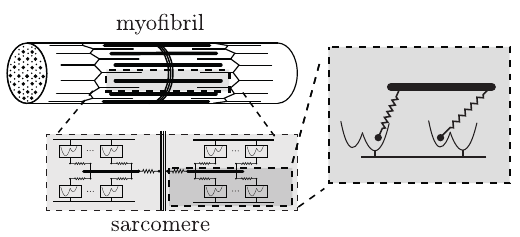}
\caption{Half-sarcomere modeled as a parallel cluster of   Huxley--Simmons units.}
\label{fig:HSSR}
\end{figure}

To rationalize the effect of the destabilizing short-range interactions, we use the general framework of the HS model. However, we now complement the mean-field interaction,  favoring uniformity of micro-configurations, by the short-range interactions favoring the inter-mixing of pre- and post-power stroke states, see Fig.\ref{fig:HSSR}. More specifically, we make the interaction of antiferromagnetic type between neighboring spins compete with the long-range paramagnetic-type interaction introduced in the original HS model. The ensuing 1D hard-spin model of a muscle half-sarcomere can be viewed as a version of the corresponding  Ising model where various short- and long-range interactions are allowed to compete. Such competition can generate a rich repertoire of thermodynamic behaviors  \cite{Nagle_PRA_1970, Nagle_JAP_1971, Kardar_PRB_1983, Nishino_PRB_2016, Campa_2019} and the goal of this paper is to explore different alternatives in the context of muscle mechanics. An additional complication is that the HS mean-field interaction makes the ensuing systems non-additive, which leads to ensemble inequivalence.  In particular, the isometric and the isotonic mechanical responses will be necessarily different and would have to be treated separately \cite{Campa_2019, Mukamel_PhysicaA_2010, Barre_PRL_2001,  Barre2002}.  While the detailed nature of the postulated antiferromagnetic interactions still remains debatable (see Sect.~2 for details), we take in this paper a position that the quantitative consequences of the new hypothesis must be explored before any judgment of its relevance for the theory muscle contractions can be made.

We show that if the antiferromagnetic short-range interactions are sufficiently strong, the HS-type coherent single-state  response  can be indeed  compromised. It is replaced by a non-HS response with cross-bridges in pre- and post-power stroke configurations finely mixed. Moreover, we show that in such regimes, the cross-bridges necessarily form a regular interdigitated pattern. The macroscopically \emph{homogeneous} response of a muscle fiber is then recovered at the expense of the   individual half-sarcomeres becoming maximally \emph{inhomogeneous}.

One can say that the presence of short-range interaction is responsible for the emergence of a new energy well which stabilizes in this model the physiological state of isometric contractions.   In addition to the coherent pre- and post-power stroke configurations anticipated by the original HS theory, the augmented model, accounting for short-range interactions, predicts the stability of configurations with pre- and post-power stroke cross-bridges finely mixed. In this new 'phase', the overall stiffness of a half-sarcomere is positive, which suggests that it may be adequately representing the physiological state of isometric contractions.   The macroscopically homogeneous response is then stabilized passively,  in contradistinction with active stabilization studied in \cite{SRT_PRE_2016}.

We observe that additional energy wells (chemical states), different from the ones describing pre- and post-power stroke conformations, are sometimes introduced phenomenologically in chemo-mechanical models to ensure that the $T_2(\delta \ell)$ curve is sufficiently flat around the stall state, see, for instance,  \cite{Marcucci_PLOS_2016} and the references cited therein. Instead, in the model with short-range antiferromagnetic interactions, the third energy well emerges directly from a micromodel without additional phenomenological assumptions.

The three-state systems of this type, which would be usually represented by a Potts model \cite{Ostilli2020, yeomans1992statistical}, are characterized by a tricritical point separating the line of first- and second-order transitions. In our two-well microscopic model,  where the third well appears after statistical averaging, the  \emph{tricritical} point arises as well,  strongly influencing the corresponding equilibrium phase diagram.  In addition to the 'coherent' bi-stable phase predicted by the original HS theory and the 'melted' phase appearing at high temperatures, such a phase diagram shows the existence of a domain with configurations of the antiferromagnetic type containing both pre- and post-power stroke cross-bridges. Acquiring such configurations may be physiologically beneficial for the system, which can then \emph{anticipate} the necessity of either ultrafast contraction  (global switching to post-power stroke state) or ultrafast stretching (global switching to pre-power stroke state). Moreover,  we argue that a living system may benefit from being posed near such a singular point.

The paper is organized as follows. Section 2 discusses the feasible mechanisms behind the antiferromagnetic short-range interactions affecting neighboring myosin heads. The augmented HS  model of a half-sarcomere is introduced in Section 3, where we study its behavior at zero temperature. Section 4 considers finite temperature effects and constructs the equilibrium phase diagram for the system in a hard device.  The system in a soft device is studied in Section 5,  where we also discuss the origin of the ensemble non-equivalence. We then specify in Section 6 the order parameters and study spatial correlations. The case study of two half-sarcomeres in series is presented in Section 7.  Finally, in Section 8, we summarize our results.

\section{Short-range interactions between neighboring myosin heads}

\begin{figure}[h!]
\centering
\includegraphics[scale=1]{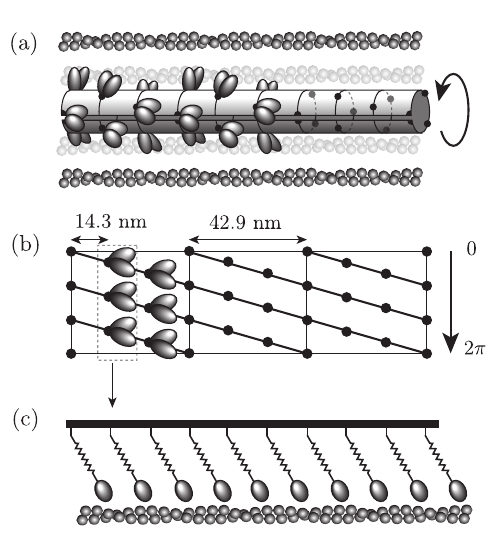}
\caption{(a) Arrangement of the myosin crowns on the cylindrical surface of the myosin backbone. (b)  Coaxial helices on flattened cylinder with   emphasized pairs of myosin heads. (c) HS stylized model   of a half-sarcomere with each cross-bridge effectively representing    three myosin crowns. The corresponding actin backbone is an effective representation of the six actin  (thin) filaments   comprising the matching stereotypical  units for each thick filament.} \label{fig:twoheads}
\end{figure}

We now turn to the physiological motivation for the introduction of the \emph{interacting heads} hypothesis. Recall first that 150 nm long myosin II molecules can be viewed as narrow rods, each ending with two  10-15 nm globular domains (heads). The heads form projections and play the role of the enzymatic active sites, the ATPases, which are activated by six adjacent actin filaments to produce muscular force and movement. Instead, the myosin tails self-associate to form the backbones of thick filaments.  The resulting rod packing is parallel with each crown (myosin head pairs) closely located to (and therefore elastically \emph{interacting} with)  several tails of the \emph{neighboring} myosin molecules. The myosin II filaments arrangement is such that the crowns form 2D  helical lattices.  The individual helices are arranged with a 14.3-nm separation between crowns and a 42.9-nm helical repeat \cite{Squire2017}, see Fig. \ref{fig:twoheads}(a).

Since the conformational change inside myosin heads produces an extension of the order of   $\sim 10$ nm \cite{Higushi_Nature2017}, the size of the working stroke is comparable to the distance between consecutive myosin heads along the helix.  Therefore, whether a particular myosin head is in pre- or post-power stroke position may influence its neighbors' state,  and the local patterning may be at least partially guided by such interaction. Depending on their spatial configurations, the two successive crowns can potentially interact either directly (sterically) or indirectly through the adjacent tropomyosin--troponin regulatory units  \cite{Razumova_BiophysJournal_2000}  or through other environmental proteins \cite{Squire2009}.

The short-range interaction between the crowns may involve first and second heads. It is known, for instance, that the regularly organized helical myosin head structures can be thought as stabilized by head-to-head interactions, however, the role of the second head in the dimeric myosin II molecule remains enigmatic. For instance, there is a range of experimental results suggesting negative cooperativity between the two heads, which implies that binding of one head to actin inhibits binding of the other \cite{Mansson2015}. The two heads may then form an asymmetric structure with the overall dimension along the helix comparable with the distance between neighboring crowns \cite{Tanner_PLOS_2007}. Thus, if one of the heads in the contracting state is oriented almost perpendicular to the fiber axis, the partner head would be stretched axially \cite{Oshima_2007}.

In Fig.~\ref{fig:twoheads}(a,b),  we show the schematic  3D configuration of the crowns and illustrate the model reduction approach employed in the derivation of the simplest HS model.  The reduced model shown in Fig.  \ref{fig:twoheads}(c) introduces an  effective  cross-bridge as collective  representation of several laterally separated crowns. Each  'collective'  myosin  head  is interacting with a similarly  schematized  actin filament, representing six separate real active filaments.  Such model neglects the presence of two myosin heads and  does not resolve the actual spatial configuration of the neighboring crowns. While  this  model accounts for the elastic interaction between the effective cross-bridges  and the  rigid  filaments, it clearly  underestimates the possibility of the geometrically allowed short-range interaction  between the neighboring effective cross-bridges.

Spatial configuration  of contracting half-sarcomeres  can be also  influenced by the elastic coupling between myosin heads  along the compliant myofilament lattice  \cite{Hill_PNAS_1980, Piazzesi1995, Hancock1997, Razumova_BiophysJournal_2000, Rice_BioPhysJournal_2003, Mijailovich_BiophysJournal_1996}. The presence of  such  elastic interaction would imply that cross-bridges do not operate independently while generating force \cite{Daniel_Biophys_1998, Chase2004, Campbell2006}. In \cite{Ford1981}, a \emph{continuum} extension of the HS  model was proposed in which the   compliance of  the thick and thin filaments was actually taken into account. This model was further developed in \cite{Mijailovich_BiophysJournal_1996}, where the authors explicitly   argued  that  "the local behavior of one myosin head must depend on the of \emph{neighboring} attachment sites."  The effects of stochasticity in this model were studied in \cite{Campbell2006}.  

\begin{figure}[ht]
\centering
\includegraphics[scale=1]{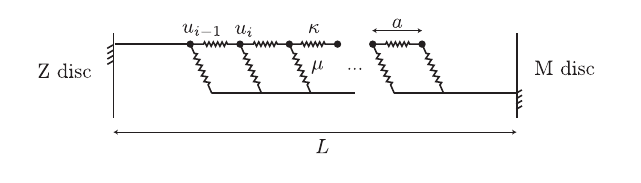}
\caption{ A schematic mechanical model of a thick  filament as an array of interacting linear springs coupled to a Winkler foundation. The cross-bridges are represented by linear springs with common elasticity $\mu$. Each cross-bridge interacts elastically with each of its neighbors via a linear  spring with  modulus $\kappa$.}\label{fig:ReaslisticSarcomere}
\end{figure}

More recently, a \emph{discrete} model dealing with  an array of motors and also accounting for   compliance of discretized myofilaments was considered in \cite{Powers2020}.  A  minimal representation of this  discrete  model of a half-sarcomere,  where   $N$  effective cross-bridges are protruding   periodically from an elastic backbone  and are bound to a rigid actin filament, is shown in Fig.~\ref{fig:ReaslisticSarcomere}. In mechanical terms, it can be viewed  a mass--spring chain containing $N$ nodes that are linked to an elastic (Winkler) foundation. Neglecting the conformational  bi-stability, one can assume  that springs are  linear with the  reference length $a = L/N$, where $L$ is the length scale of a half-sarcomere.  Such  model was probably first introduced  by  De Gennes in his  pioneering study of the mechanical response of  two-stranded DNA \cite{DEGENNES20011505}.  

To assess the nature  of the effective nearest neighbor (NN) interaction between the nodes, we can  reduce the 'double chain' model shown in Fig.~\ref{fig:ReaslisticSarcomere} to a 'single chain' minimal model. To this end, we introduce the displacements of the nodes  $u_i$  and write the  energy of the system in the form 
\begin{equation}\label{1}
E(\boldsymbol{u}) = \frac{1}{N}\sum_{i=1}^N \frac{\kappa}{2}\left(\frac{u_{i}-u_{i-1}}{a}\right)^2 +\frac{1}{N}\sum_{i=1}^{N+1} \frac{\mu}{2}u_i^2, 
\end{equation}
where  $\kappa$ is the stiffness of the horizontal springs and  $\mu$ is the (shear) modulus of the vertical (leaf) springs. While the first term in Eq.~\eqref{1} describes the compliance of the  filaments, the second term represents the elasticity of the individual cross-bridges. We note that for simplicity,  the thick filament is assumed to be rigid. 

In the limit $a \to 0$ and $N \to \infty$, the discrete model  \eqref{1} can be  approximated by the continuum model with energy
\begin{equation}
\label{2}
E(u) = \int_0^L \left( \frac{\kappa}{2}  w^2 + \frac{\mu}{2} u^2\right) dx,
\end{equation}
where $x\in(0,L)$ is the spatial coordinate,  $u(x)$ is the horizontal displacement field, and $w(x)= \partial_x u(x)$ is the continuum strain field. We can assume that the system is subjected to isometric loading (hard device) with the total displacement prescribed  $\int_0^L w dx = d$. 

To rewrite the energy \eqref{2} in terms of the local strain field $w$ only, we can follow \cite{Ren2000} and   define the characteristic (indicator) function $\xi_y:x \to \{0,1\} $ such that for $x\leq y$ : $\xi_y(x) = 1$, and for $x> y$ : $\xi_y(x) = 0$. If we  then assume for determinacy that $u(0)=-d/2$, and $u(L) = d/2$, we can write
\begin{equation}
\begin{split}
\int_0^L \mu u^2 dx & =
  \mu \int_0^L \left[\int_0^L \xi_x(y) w(y) dy -\frac{d}{2} \right]^2 dx 
\\
& = 
\mu \int_0^L \left[\int_0^L \left(\xi_x(y) w(y)  -\frac{1}{2}w(y)\right)dy \right]^2 dx 
\\
\end{split}
\end{equation}
Then, noticing that 
$\int_0^L\left(\xi_x(y)  -\frac{1}{2}\right)\left(\xi_x(z)  -\frac{1}{2}\right)dx = \frac{L}{4}-\max\{x,y\}+\frac{1}{2}(x+y),$
we can rewrite the energy \eqref{2}  in the form
\begin{equation}
\label{3}  
E(w) = \int_0^L  \kappa w^2  dx + \mu \int_0^L\int_0^L K(x,y)w(x)w(y)dx dy
\end{equation}
where $K(x,y) = \frac{L}{4}-\max\{x,y\}+\frac{1}{2}(x+y)$ is the interaction kernel, which is defined on the square domain $[0,L]\times [0,L]$. 

\begin{figure}[h!]
\centering
\includegraphics[scale=.9]{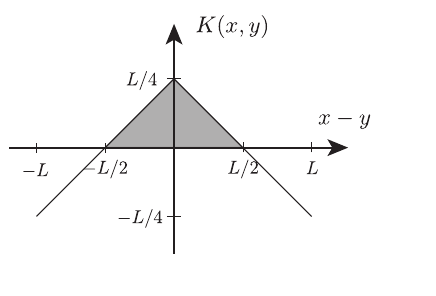}
\caption{The dependence of the kernel function $K(x,y)$ on  $x-y$. For $|x-y|\leq L/2$, we observe antiferromagnetic-type behavior with $K\geq0$. }\label{fig:Kernel}
\end{figure}

In Fig.~\ref{fig:Kernel}, we illustrate the structure of the kernel function $K(x,y)$ showing its behavior along the  $x-y$ diagonal direction.  Note that  $K(x,y)>0$ for small values of  $|x-y|$, which is the range corresponding to   short-range interactions. Therefore,  the configurations with $w(x)$ and $w(y)$ having  alternating signs  at neighboring points $x$ and $y$  will have lower energy   $E$ vis a vis the configurations with locally homogeneous strain. As a result,   the system with the energy \eqref{3}  will be driven toward configurations with locally alternating strains ($w>0$  for stretching and $w<0$  for compression).  Such configurations can be  interpreted as  having   \emph{antiferromagnetic order} \cite{chaikin1995principles}.
In the presence of nonlinear terms in the energy describing conformational change (power stroke), the underlying effective antiferromagnetic-type short-range elastic interaction between neighboring nods (describing myosin heads)  can be expected to facilitate the formation of fine \emph{mixtures}  of cross-bridges in pre- and post-power stroke  configurations.

We can conclude that the account of steric effects and filament extensibility can, in principle, bias the relative displacements of the neighboring cross-bridges toward non-uniformity \cite{Mijailovich_BiophysJournal_1996, Daniel_Biophys_1998}. Such interactions make the local conformational state of an attached myosin head affected by the conformational state of the neighboring heads. Moreover, it is known \cite{Ren2000} that the account of the antiferromagnetic interaction can make the coherent configurations with larger averaged spacing energetically less preferable than the 'closed packed' interdigitated configurations with pre- and post-power stroke cross-bridges periodically patterned at the smallest available scale.

Consider now the issue of conformational homogeneity in the mechanical response of single half-sarcomeres. It is well known that in isometrically contracting muscle myofibrils, the cross-bridges move through the reversible power stroke asynchronously and assume, at any moment in time and any given location, \emph{both} conformations \cite{MIDDE20111024, Hirose1993, Reedy_2000, Thomas_BiophysicalJornal_1995}. In particular, there is evidence that at the steady-state plateau of isometric tetanus, see Fig. \ref{fig:Isometric}(a), the individual myosin motors are in different conformations, which is sometimes interpreted as a 'state of disorder.' However, the cross-bridges can potentially get synchronized spatially and temporarily in response to very rapid stretches or slacks and, in this sense, the fast force recovery is sometimes interpreted,  in accordance with HS theory, as a disorder-to-order transition \cite{Thomas_BiophysicalJornal_1995}.  

In any case, such interpretations remain qualitative since the usual mechanical and structural measurements at the single half-sarcomere level address populations of myosin motors. This makes it difficult to follow the change in conformation of individual cross-bridges and quantify the process of stress-induced spatial synchronization \cite{Lombardikappa}. Still, specially designed X-ray experiments can provide at least some access to the spatial distribution of the myosin motors. For instance, one can use for this purpose the width of the third-order myosin-based meridional reflection (M3 reflection) as it can be related to the number of myosin motors in the pre-power stroke state.  Moreover, the splitting of the M3 reflection can be linked to the distance between the two arrays of myosin motors in pre- and post- power stroke states \cite{Reconditi_PNAS_2017}. Based on these and other experimental approaches, it was suggested that while at loads closer to stall conditions myosin motors are broadly distributed, at lower loads they proceed through the stroke almost collectively  \cite{Piazzesi1992, Reconditi_Nature_2004}. 

Such measurements, however, still provide information only about the homogenized state. If,  for instance,  the averaged observed state in the quenched configuration is between pre- and post-power stroke, they do not reveal the actual \emph{geometrical (ordered) pattern} behind this averaged behavior.

To summarize,  the assumption of the original HS model that individual myosin heads interact only through the thick filament,  modeled as a rigid backbone, does not allow one to make any conclusions about the spatial organization of pre- and post-power stroke cross-bridges inside a single half-sarcomere.  To quantify the conjectured self-organization of the closely interacting myosin heads into alternating conformational patterns, it is necessary to go beyond the class of HS type mean-field models and account explicitly for short-range interactions between neighboring cross-bridges.  Such in silico modeling will allow one to make quantitative predictions which are needed to either confirm or reject the hypothesis of antiferromagnetic-type nearest neighbor interactions.

\section{The  model of a half-sarcomere}

Following HS, we model a half-sarcomere as a collection of $N$ interacting cross-bridges, but we supplement the mean-field interaction introduced of HS by a competing short-range interaction. 

\begin{figure}[t]
\centering
\includegraphics[scale=1]{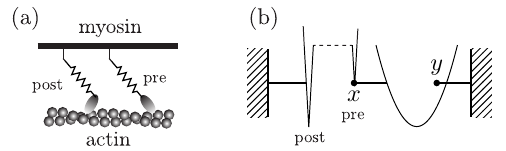}
\caption{ (a) Schematics showing two neighboring cross-bridges in different conformational states, (b)  Huxley--Simmons (HS) model of a single cross-bridge.  The two conformational states of the myosin head are represented by the spin variable $x$, while the lump elasticity of the whole cross-bridge is represented by the parabolic potential acting on the continuous variable $y$.}
\label{fig:CB}
\end{figure}

The energy of  non-interacting cross-bridges can be written as  $$\sum_{i}^{N}V(X_i), $$ where $i=1,\dots, N$ and $ V(X_i) $ is a  double-well energy of a single cross-bridge, see Fig.~\ref{fig:HSSR}. We assume for analytical simplicity that the two conformational positions of the myosin head are represented by the hard-spin variable $X_i$ so that in the pre-power stroke state $X_i=0$, and in the post-power stroke state  $X_i=-a$, where $a$ is the amount by which the myosin head pulls the actin during the power stroke. We set that $V(X_i)=(a+X_i)V_0$ which shows that the pre-power stroke conformation has higher energy than the post-power stroke conformation with the energy bias equal to $V_0$ (in the units of force), see Fig.~\ref{fig:CB}. Despite the overall passive (ATP indifferent) nature of the HS model, the parameter $V_0$ has an active origin as  it is ultimately responsible for the  generation of the  tension $T_0$ in the stall state (state of isometric contractions).  

In the HS model, each cross-bridge is connected to a  common rigid backbone through an elastic spring $\kappa_0$ placed in series with a bi-stable unit. The backbone is characterized by a single variable $Y$, and the corresponding interaction energy is of mean-field type $$\sum_{i}^{N} \frac{\kappa_0}{2}(Y-X_i)^2, $$ where $\kappa_0>0$ is a parameter stabilizing homogeneous distribution of cross-bridge configurations.

In addition to this interaction, we assume that each element also interacts harmonically with its nearest neighbors through the energy term  $$\sum_{i}^{N} \frac{\kappa_J}{2}(X_{i+1}-X_i)^2,$$ where  $\kappa_J$ is the corresponding   linear stiffness. The various potential sources of such  short-range interaction were discussed in detail in Sect. 2. In particular, we  showed there that   to adequately account for compliance of the filaments and to produce  the anticipated  destabilizing effect on the homogeneous distribution of cross-bridges, such short-range  interaction should be of antiferromagnetic type. This means that the parameter $\kappa_J$ should   be negative.
 
Finally, we assume that the whole half-sarcomere is loaded through an elastic spring with stiffness $\kappa_f>0$,   representing the combined elasticities of actin and myosin filaments. When the system is loaded in a hard device, an imposed displacement $Z$ is applied to the external spring giving the contribution to the energy $$\sum_{i}^{N} \frac{\kappa_f}{2N}(Z-Y)^2.$$


To non-dimensionalize the problem, we define the reference length $a$ and normalize the spatial variables accordingly: $x_i=X_i/a$, $y=Y/a$ and $z=Z/a$.  Note that now  the variable $x_i$ takes values 0 and -1 for the pre- and post-power stroke, respectively. Using $\kappa_0 a^2$ as the scale of  the  energy, we obtain in  dimensionless variables
\begin{equation}\label{eq:enehdsr}
\begin{split}
E(\bm{x},y,z)&=\sum_{i}^{N}\left[(1+x_i)v_0+\frac{1}{2}(y-x_i)^2 +\frac{\lambda_J}{2}(x_{i+1}-x_i)^2+\frac{\lambda_f}{2}(z-y)^2\right],
\end{split}
\end{equation}
where $v_0 = V_0/(\kappa_0 a)$ is the non-dimensional energy bias. We consider periodic boundary conditions:  $x_{N+1}=x_1$, however,  in the thermodynamic limit $N \to \infty$; the choice of this type of boundary conditions  becomes irrelevant \cite{yeomans1992statistical, goldenfeld1992lectures}.  The ensuing problem  contains two dimensionless parameters  $\lambda_J = \kappa_J/\kappa_0$ and   $\lambda_f = \kappa_f/(N\kappa_0)$. 
 
To reveal the main effect, it is sufficient to analyze the system at zero temperature. To obtain the ground state, we first eliminate $y$ using the condition $ \partial E/\partial y=0$, which is equivalent to 
\begin{equation}\label{eq:y_equilibrated}
y=\frac{\lambda_f z}{1+\lambda_f}+\frac{1}{N(1+\lambda_f)}\sum_i x_i.
\end{equation}
The relaxed energy is of Ising form with both mean-field and short-range interactions: 
\begin{equation} \label{eq:Hamiltonian}
\begin{split}
E(x_i,z) =-\frac{K}{2N} \sum_{i,j} x_i x_j - \lambda_J\sum_i x_{i}x_{i+1}-h(z) \sum_i  x_i +c(z).
\end{split}
\end{equation}
where $1/K= 1+\lambda_f$  and  $$h(z) = \frac{\lambda_f z}{1+\lambda_f} - v_0 +\lambda_J+\frac{1}{2},\,\,\,\, c(z) =  \frac{N\lambda_fz^2}{2(1+\lambda_f)} +N v_0.$$ Since this energy is quadratic in $z$ at a given $x_i$, the minimization over  $x_i$ reduces to the choice among a finite number of parabolas.
%

Depending on the values of $\lambda_J$ and $\lambda_f$, the global minima correspond to either pure phases, with all cross-bridges synchronized in the same conformational state (pre- or post-power stroke), or to a mixed phase with lattice-scale oscillations between the two conformational states. At a not too negative value of $\lambda_J$, the energy minimizing phase is ferromagnetic with pre- or post-power stroke state dominating depending on whether the system is stretched or compressed. At sufficiently negative $\lambda_J$, when destabilizing short-range interactions are strong, the unstressed system can also be in an antiferromagnetic state when neighboring cross-bridges are in opposite conformational states.

\begin{figure}[ht!]
\centering
\includegraphics[scale =1 ]{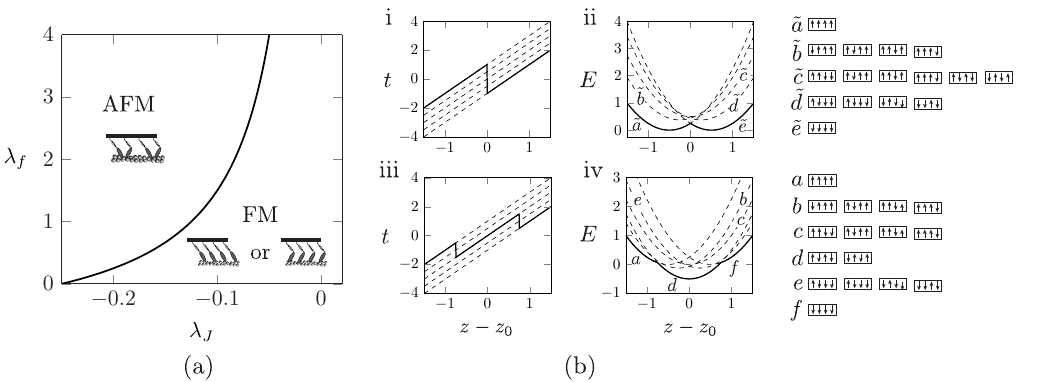}
\caption{ (a) Zero temperature phase diagram for a half-sarcomere in the stall state $z=z_0$. (b) Energy and tension of a system of $N= 4$ cross-bridges. The thick solid line represents the global minimum of the energy (ground state),  while the dashed line represents the local energy minima (metastable states). In i--ii, we illustrate the ferromagnetic (FM)  part of the phase diagram (corresponding to $\lambda_f =1$, $\lambda_J = 0$), and in iii--iv,   the antiferromagnetic (AFM) part (corresponding to $\lambda_f =1$, $\lambda_J = -0.5$). In (b), each metastable branch is associated with a particular micro-configuration, which is illustrated in the right column. The 'up/down arrows represent  pre/post power stroke states. }
\label{fig:groundstate}
\end{figure}

The phase diagram, revealing the structure of the ground state in the unstressed state, is presented in Fig.~\ref{fig:groundstate}(a). The boundary separates the antiferromagnetic (mixed) and ferromagnetic (pure)  ground states. We assumed here, for simplicity, that the unstressed state describes the stall condition and also arbitrarily set that in this state, the distribution of pre- and post-power strokes cross-bridges is 50-50 \footnote{While it would have been more realistic to assume that this ratio is around 70-30 \cite{Irving2000}, we have chosen to present our results only in the most simple setting that can be easily corrected by the appropriate adjustment of the parameter $v_0$.}.
We therefore have   the stall state at  $z = z_0=(1+1/\lambda_f) v_0-1/2$  where   post- and  pre-power stroke  conformations have the same energy. Note that at $z=z_0$ the  energies of pure and mixed configurations become equal if 
$
1/\lambda_J = - 4(1+\lambda_f).
$
At $$\lambda_J < -\frac{1}{4(1+\lambda_f)}$$ the equilibrium configuration in stall conditions  is  mixed,  while  it is coherent  (all cross-bridges are either pre- or post-power stroke) otherwise. In the limiting case $\lambda_f \to \infty$, when cross-bridges  interact only  at  short-range,     the transition  occurs at $\lambda_J = 0$. In the other limiting case  $\lambda_f \to 0$, the crossover between ferromagnetic and antiferromagnetic  phases takes place  at $\lambda_J = -1/4$.

We now compare the energies of different spatial distributions  $x_i$  at $z =z_0$. In the FM  phase, any of the mixed configurations will have higher energy than one of the  pure configurations $E_{mixed}(z = z_0) > E_{pure}(z = z_0)$, while in the  AFM phase $E_{mixed} (z = z_0) < E_{pure}(z = z_0)$. Here $E_{pure}$ can be taken as   the energy of a configuration with either  all pre- or all post-power stroke  cross-bridges because  at $z = z_0$ they are equal.


In Fig.~\ref{fig:groundstate}(b)(ii, iv), we show the energies of local and global minimizers $E(z)$ corresponding to mixed (alternating pre- and post-power stroke states) and pure (fully synchronized) configurations that are parameterized by $z$. The corresponding branches of tension--elongation curves  $t = d E/dz$   are shown in Fig.~\ref{fig:groundstate}(b)(i,iii).

It is interesting that in the tension curves \ref{fig:groundstate}(b)(i, iii), different metastable branches are uniquely characterized by the total amount of spins in either pre- or post-power stroke, which means that from developed tension alone we cannot distinguish spatial organization. To this end, we need to turn to the energy function. The reason comes directly from the energy \eqref{eq:Hamiltonian} where all $z$ related terms are in $h$ and $c$, which means that both short- and long-range terms ($J$ and $\lambda_J$, respectively) are $z$ insensitive, and only the average 'magnetization' $\sum_i x_i$ interacts with the loading device. However, ultimately the choice of the most stable branch is dictated by the energy.

We also note that in the ferromagnetic phase, the macroscopic stiffness in the stall state is equal to minus infinity. In the antiferromagnetic phase, the transition between the two pure states is smoothed, but the stiffness in the stall state remains to be equal to minus infinity. This means that the formal 'stabilization' of this state is impossible at zero temperature.

To summarize, the analysis of the zero temperature system shows that due to the mean-field nature of ferromagnetic interactions, the preferred behavior is coherent with fully relaxed energy remaining non-convex and the corresponding stiffness taking negative values. Adding sufficiently strong short-range antiferromagnetic interactions partially convexifies the relaxed energy; however, negative stiffness persists.

\section{Equilibrium response }

To study the equilibrium response of the system at finite temperature $\theta$, we need to compute its free energy. The  partition function in the \emph{hard device} ensemble (controlled displacement) is,
$$\mathcal{Z}(\beta,z) = \int dy\sum_{\{x\}}e^{-\beta E(x_i,y,z)}, $$
where the summation is over $x_i=\{0,-1\}$ for all $x_i$; we also use the standard notation  $\beta=\kappa_0 a^2 /k_B \theta$, where  $k_B$ is the Boltzmann constant.    Since $\sum_i x_i^2=-\sum_i x_i$, we can write   
\begin{equation}\label{eq:partfuncsr}
\begin{split}
\mathcal{Z}(\beta,z) &= \int dy e^{-\beta N \varphi(y,z)} \mathcal{Z}_0(\beta,y),
\end{split}
\end{equation}
where  $\varphi(y,z)=\frac{\lambda_f}{2}(z-y)^2+v_0+\frac{y^2}{2}$ and  the function $\mathcal{Z}_0(\beta,y)$ is the partition function for an Ising ring \cite{goldenfeld1992lectures}: 
\begin{equation}\label{eq:Z0}
\begin{split}
\mathcal{Z}_0(\beta,y)&=\sum_{\{x\}}\exp\left[\beta J \sum_{i}x_i x_{i+1}+\beta H \sum_i x_i  \right],
\end{split}
\end{equation}
where, $J=\lambda_J$ and $H(y)= \lambda_J+y+1/2-v_0$.  The equilibrium  behavior  of this system without short-range interactions ($\lambda_J=0$) was studied in \cite{CT_PRE}. 
\begin{figure}[ht!]
\centering
\includegraphics[scale=.9]{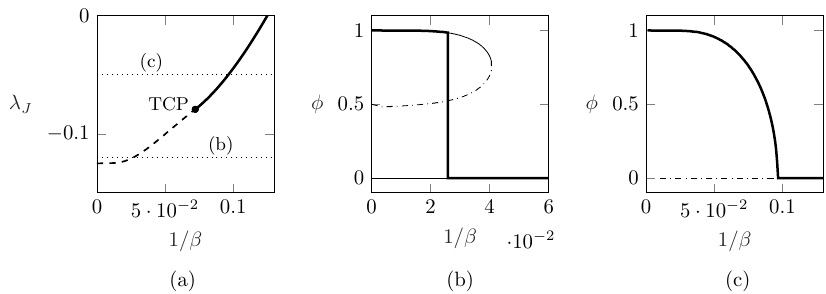}
\caption{Different solutions of the self-consistency condition \eqref{eq:scsr} correspond to interactions of different curves with an inclined straight line. In Phase I, there are three solutions (two energy minima, two symmetric ground states), in Phases II and III, there are five solutions (three energy minima with either one or two ( symmetric) ground states),  and in Phase IV, the solution is unique (single energy minimum which is also the ground state).}
\label{fig:SelfConsistSR}
\end{figure}

\begin{figure}[ht!]
\centering
\includegraphics[scale = 1]{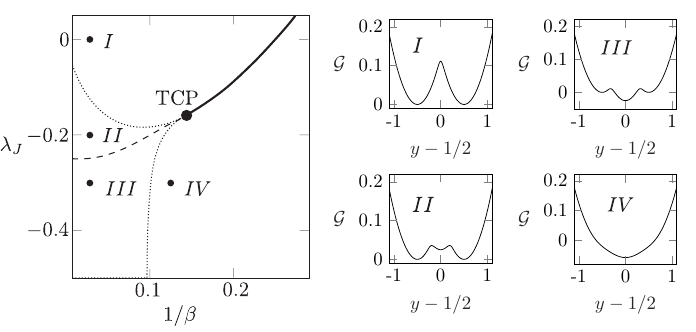}
\caption{  (a) Phase diagram for the system in the hard device with $z=z_0$. Parameters: $\lambda_f = 1$, $v_0=0$. The transition between Phases I and IV is continuous (bold solid line) down to the tricritical point TCP, where it becomes first-order (dashed line). The dotted lines are spinodal boundaries. The different phases are illustrated in the right column, where we show the dependence of the partial free energy $\mathcal{F}$ on the internal variable $y$.}
\label{fig:PhaseDiagram}
\end{figure}

To compute $\mathcal{Z}_0(\beta,y)$ we can use the  transfer matrix method  \cite{Campa_2019,christensen2005complexity}. The idea is to represent this partition function as a product of equal matrices \cite{yeomans1992statistical, goldenfeld1992lectures}. Consider the   matrix 
\begin{equation}\label{eq:Tmatrix}
\bm{T}=
 \left(\begin{array}{cc}
1 & e^{-\frac{\beta H}{2}} \\ 
e^{-\frac{\beta H}{2}} & e^{\beta (J-H)}
\end{array} \right).
\end{equation} 
Then,
$
\mathcal{Z}_0=
 \Tr(\bm{T}^N),
$
and since $\bm{T}$ is real and symmetric, it can be diagonalized with the eigenvalues 


\begin{equation}\label{eq:eigenvalues}
\lambda_{1,2}=e^{-h}\left[ \cosh h \pm\sqrt{\sinh^2h+e^{-\beta J}} \right],
\end{equation}
where $h =\frac{\beta}{2}(H-J)$. Using the invariance of the trace, we obtain 
$
\mathcal{Z}_0=\lambda_1^N+\lambda_2^N.
$
If we now assume that $\lambda_1>\lambda_2$ and 
take  the thermodynamic limit $N\rightarrow \infty$, we obtain 
$
\mathcal{Z}_0\approx\lambda_1^N\left[1+\mathcal{O}(e^{-\alpha N})\right],
$
where $\alpha \equiv \log(\lambda_1/\lambda_2)$ is a positive constant. 
Using the Laplace method, we finally obtain
\begin{equation}
\mathcal{Z}(\beta,z) \sim \min_y \left\lbrace e^{-\beta \varphi (y, z)}  \lambda_1 (y) \right\rbrace.
\end{equation}
The free energy density  is $\mathcal{F}(\beta,z)=-\frac{1}{N\beta}\log \mathcal{Z}(\beta,z)$.   In the thermodynamic limit,  we can write 
\begin{equation}\label{eq:FE}
\begin{split}
\mathcal{F}(\beta,z,y)&=   \frac{\lambda_f}{2}(z-y )^2+v_0+\frac{y ^2}{2}+\frac{g(y )}{\beta}
 -\frac{1}{\beta}\log\left[ \cosh g(y )+  \sqrt{e^{-\beta \lambda_J}+\sinh^2 g(y )} \right].
\end{split}
\end{equation}
where  $g(y) =  \frac{\beta}{2} (y+1/2-v_0)$.  By solving a transcendental equation $ \partial \mathcal{F}(\beta,z,y)/\partial y=0$, we obtain
 the  self-consistency relation 
 $y_* = \Psi (\beta, z, y_*)$ illustrated in Fig.~\ref{fig:SelfConsistSR} with the function $\Psi (y_*)$ known  explicitly  
\begin{equation}\label{eq:scsr}
\Psi (\beta, z, y )=\lambda_f (z- y )-\frac{1}{2}+\frac{1}{2}\frac{ \sqrt{e^{-\beta  \lambda_J}+\sinh ^2 g(y )}}{ e^{-\beta  \lambda_J} \csch g(y )+ \sinh g(y )}.
\end{equation}
As we show in Fig.~\ref{fig:SelfConsistSR}, the equation \eqref{eq:scsr} may have more than one solution, which reflects  the non-convexity of the partial free energy    $\mathcal{F}(\beta,z,y)$
The parametric dependence of this energy is illustrated  in Fig.~\ref{fig:PhaseDiagram} for the stall state with $z=z_0$. Four different phases  can be identified  on the  plane  ($\lambda_J, 1/\beta$). The zero temperature phase diagram shown in Fig.~\ref{fig:groundstate}(a) corresponds to the first-order transition boundary separating Phases II and III and  is represented in Fig.~\ref{fig:PhaseDiagram} by its section at  $\lambda_f = 1$. 

In the domain with low temperature and dominating mean-field interactions, we observe Phase I.  In this phase, two pure states coexist, but at any instant of time, all cross-bridges are either in pre- or post-power stroke conformations. In other words, in stall conditions, the system in Phase I randomly switches between these two states. Each of these transitions takes place coherently, with all cross-bridges changing their state cooperatively.  In the high-temperature domain and still dominating mean-field interactions, we observe Phase IV, where all cross-bridges switch randomly and independently between  pre- or post-power stroke conformations. Phase IV is separated from Phase I by a second-order phase transition, analogous to the usual order--disorder transition in magnetics.

At low temperatures and strong short-range antiferromagnetic interactions, we observe the appearance of Phases II and III, where in addition to pure states, there is also a mixed state where neighboring cross-bridges have different conformations.  In Phase II, the mixed state is metastable, and the equilibrium states are the pure ones, while in Phase III, the equilibrium state is the mixed one. Phases II and III are separated by the line of first-order phase transition, which meets the second-order phase transitions line separating Phases I and III at a tricritical point. 


To summarize, when the short-range interactions are sufficiently strong and the temperature is sufficiently low, we observe a new feature in the system's behavior: An antiferromagnetic state becomes stable in the stall regime $z=z_0$ (our Phase III).  In other low-temperature Phases I and II, the stall state must necessarily involve random coherent conformational changes involving all cross-bridges simultaneously. At high temperatures, the paramagnetic phase dominates where all cross-bridges fluctuate randomly and independently over the whole energy landscape with neither pre- or post-power stroke stated clearly discernible. We have seen that by increasing  $\lambda_J$ in the positive direction, we effectively decrease the temperature by favoring cooperativity and effectively destabilizing the homogeneous state at $z=z_0$. Instead, by increasing  $\lambda_J$  in the negative direction, we disfavor coherent fluctuations between pure states at $z=z_0$ and stabilize instead non-fluctuating macroscopically homogeneous state, which we later show to be microscopically a fine lattice-scale mixture of pre- and post-power stroke conformations.



\begin{figure}[ht!]
\centering
\includegraphics[scale=1.1]{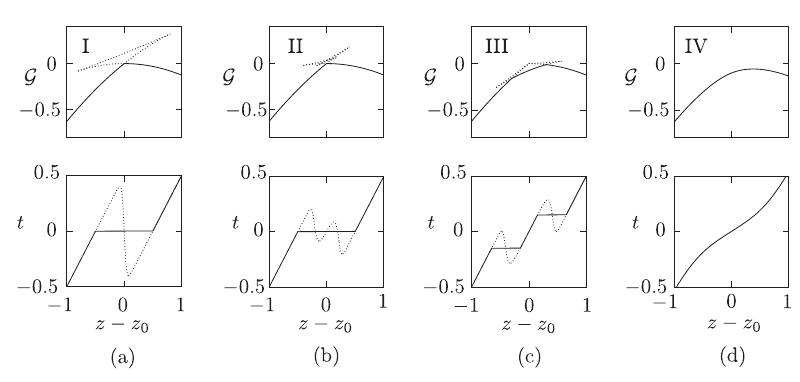}
\caption{ The equilibrium free (Helmholtz) energy--strain relation in the hard device and the corresponding  tension--strain relation.  (a) Phase I: $\lambda_J= 0$, $\beta = 50$; (b) Phase II: $\lambda_J = - 0.1$, $\beta = 50$; (c) Phase III: $\lambda_J = -0.25$, $\beta = 50$; and (d) Phase IV: $\lambda_J = -0.25$, $\beta = 8$.  Other parameters:  $\lambda_f = 1$ and $v_0 = 0$. Solid curves correspond to ground states. With  dotted lines, we show for convenience  the unstable equilibria which have nothing to do with the equilibrium response but provide information about the energy barriers.} 
\label{fig:FE_TE_HD}
\end{figure}

We now turn to the parametric behavior of the equilibrium free energy $\mathcal{F}(z)=\mathcal{F}(z,y_*(z))$. Knowing this energy, one can also compute the equilibrium tension  $t(z)= d \mathcal{F}(z)/dz = d\mathcal{F}(z,y_*(z))/d z= \partial \mathcal{F}(z,y_*(z))/\partial z=\lambda_f(z-y_*(z)),$ where we temporarily omitted dependence on $\beta$. The behavior of the functions  $\mathcal{F}(z)$ and $t(z)$ in different phases is illustrated in Fig.~\ref{fig:FE_TE_HD}. Note that in (a) and (b) (Phases I and II), the free energy is non-convex at  $z=z_0$, and the corresponding stiffness is negative. Instead, in (c) and (d) (Phases III and IV), the stall state is associated with the point of convexity of the equilibrium energy, and the corresponding stiffness is positive. However, if in Phase IV we deal with entropic stabilization at the expense of an identifiable conformational state, in Phase III, we encounter macroscopic homogeneity with the perfect antiferromagnetic order where each cross-bridge maintains its conformation with one half of them being in pre- and another half in post-power stroke state.

To obtain an analytical expression for the line of critical points (and ultimately for the tricritical point where it ends)  on the phase diagram shown in Fig.~\ref{fig:PhaseDiagram}, we can build around it a  polynomial, Landau type expansion  \cite{ goldenfeld1992lectures, pathria1996statistical}.

To define an order parameter, we recall that, according to \eqref{eq:y_equilibrated},   the average value of the  microscopic spin variable $\mean{x}= N^{-1}\sum_i x_i$ is related to the macroscopic variable $y$ through the relation
$
\mean{x}=(\lambda_f+1)y-\lambda_f z.
$
Since $E(x_i=0,z_0)=E(x_i=-1,z_0)$, in stall conditions $z=z_0$ the  fraction of post-power stroke  elements should be equal to the fraction of pre-power stroke  elements,  which suggests that  $\left<x\right>(\beta,z_0)=-1/2$. Then, the natural order parameter is $\phi=2\mean{x}+1$, or equivalently, 
$\phi=2(\lambda_f+1)y-2\lambda_f z+1.$ It now  represents the fraction of elements in either conformation, meaning that with $\phi = 1$  all units are in the pre-power stroke, and with $\phi=-1$ in the post-power stroke configuration. Finally, $\phi= 0$ means that there are  equal number of elements in both conformations.

The marginal free energy  at $ z=z_0$ can be now written in terms of  $\phi$ 
\begin{equation}\label{eq:FEop}
\begin{split}
\mathcal{F}(\beta,\phi)=\frac{1}{2}\left(v_0+\frac{1}{2}\right)^2+\frac{v_0^2}{2\lambda_f}+\frac{\phi^2}{8(1+\lambda_f)}
-\frac{1}{\beta}\log\left[\cosh\frac{\beta\phi}{4(1+\lambda_f)}+\sqrt{e^{-\beta \lambda_J}+\sinh^2\frac{\beta\phi}{4(1+\lambda_f)}}\right].
\end{split}
\end{equation}
The corresponding self-consistency relation expressed in terms of  $ \phi_*$ takes the form 
\begin{equation}\label{eq:scsrop}
\phi_*=\frac{\sinh\frac{\beta \phi_*}{4(1+\lambda_f)}}{\sqrt{e^{-\beta \lambda_J}+\sinh^2\frac{\beta\phi_*}{4(1+\lambda_f)}}}.
\end{equation}
\begin{figure}[t]
 \centering
\includegraphics[scale = 1]{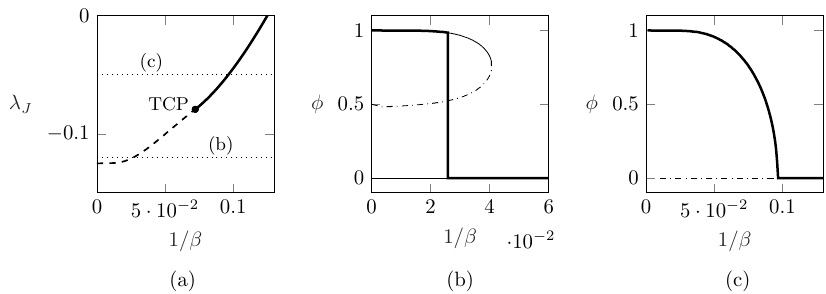}
\caption{ (a) Phase diagram in the hard device with $z=z_0$. Second-order transitions are shown by a solid bold line,  first-order phase transitions by a dashed bold line. The dotted lines indicate the locations of the two representative sections shown in (b) and (c).  (b) First-order phase transition, $\lambda_J = -0.12$. Thin solid lines correspond  to metastable states, dash-dotted line---to unstable states; (c) Second-order phase transition, $\lambda_J = -0.05$.  The dash-dotted line corresponds to unstable states.}\label{fig:Phase_Transition}
\end{figure}

In Fig.~\ref{fig:Phase_Transition} (b) and (c), we show the behavior of the function $\phi_*(\beta)$ around the lines of first- and second-order phase transitions (critical points), respectively, shown in Fig.~\ref{fig:Phase_Transition}(a). Our goal now is to capture the line of second-order phase transitions separating Phases I and IV.  To this end, it is sufficient to perform the Taylor expansion of the free-energy Eq.~\eqref{eq:FEop} around $\phi=0$. We obtain the expression 
$
\tilde{\mathcal{F}}(\beta,\phi)= a_0 + a_2 \phi^2+a_4 \phi^4$
where
\begin{subequations}\label{eq:landau_coeff}
\begin{align}
&a_0 = \frac{1}{2}\left(v_0+\frac{1}{2}\right)^2+\frac{v_0^2}{2\lambda_f}-\frac{\log \left(\sqrt{e^{-\beta  \lambda_J}}+1\right)}{\beta },
 \\
 &a_2 = \frac{1}{32(1+\lambda_f)} \left[4-\frac{\beta }{(1+\lambda_f) \sqrt{e^{-\beta  \lambda_J}}}\right],
 a_4 = \frac{\beta ^3  \left(3 e^{\beta  \lambda_J}-1\right)}{6144(1+\lambda_f)^4 \sqrt{e^{-\beta  \lambda_J}}}.
\end{align}
\end{subequations}
The approximate  self-consistency relation  $ d\tilde{\mathcal{F}} 
/d\phi=0$ gives either 
%
$\tilde{\phi}=0$ or
\begin{equation}
 \tilde{\phi}=\pm  4(1+\lambda_f) \sqrt{\frac{ 6 \beta -24 (\lambda_f+1) \sqrt{e^{-\beta  \lambda_J}}}{\beta ^3 \left(3 e^{\beta  \lambda_J }-1\right)}}.
 \end{equation}
The critical inverse temperature is implicitly given by relation $ 4(1+\lambda_f) \sqrt{e^{-\beta_c  \lambda_J}})= \beta_c$, which corresponds to $a_2 =0$.
In the limit  $\lambda_J\rightarrow 0$, we obtain  $\beta_c=4(1+\lambda_f)$. 



To locate the tricritical point (TCP) where   the second-order phase transition becomes the first-order phase transition, we need to use the sixth-order expansion in the order parameter   
$
\tilde{\mathcal{F}}(\beta,\phi)= a_0 + a_2 \phi^2+a_4 \phi^4+a_6 \phi^6
$
where the additional coefficient is 
\begin{equation}
a_6 = \frac{ \beta ^5 \left(30 e^{\beta  \lambda_J}-45 e^{2 \beta  \lambda_J}-1\right)}{2949120 \sqrt{e^{-\beta  \lambda_J}}}.
\end{equation}
The tricritical point can be linked to the vanishing of the second- and fourth-order terms in the expansion. Therefore, we obtain two equations  $\beta=4(1+\lambda_f)e^{-\beta \lambda_J/2}$ and $e^{-\beta \lambda_J/2}=\sqrt{3}$. Their solution can be found explicitly $\beta_{TCP}=4(1+\lambda_f)\sqrt{3}$ and $\lambda_{JTCP}=-\log3/4(1+\lambda_f)\sqrt{3}$. 
Finally, the location of the  line of first-order transitions (Maxwell line) can be  obtained by solving the equation $\mathcal{F}(\beta,\phi=0)=\mathcal{F}(\beta,\phi=\phi_*)$, where  
\begin{equation}
\phi_*=\pm \sqrt{\frac{-a_4+\sqrt{a_4^2-3a_2a_6}}{3a_6}}.
\end{equation}
The obtained analytical relations allow one, for instance, to estimate the minimal level of antiferromagnetic short-range interactions, which is necessary to ensure the stabilization of the macroscopically homogeneous configuration of cross-bridges in the stall state.


In our previous work \cite{CAT_PRL, BRT_PRL_2019}, we argued that the peculiarity of muscles mechanical response in the state of isometric contractions can be linked to the presence in the HS-type models of muscle contraction of a critical point located in roughly the same range of parameters. Here, we see that the appearance of an additional parameter $\lambda_J$, which scales the non-HS, short-range interactions, turns such a critical point into a critical line that ends in a tricritical point. The additional degeneracy of the system around such  TCP  can lead to new physical effects that may be physiologically advantageous and deserve a special study.

\section{ Ensemble inequivalence } 
 
Due to the presence in this system of long-range interactions, the phase diagrams corresponding to the cases of hard (isometric) and soft (isotonic) loading conditions may differ \cite{Ruffo2009}.

Suppose our half-sarcomere is loaded isotonically (in a soft device) by an applied force with dimensionless value $t=\bar{T}/\kappa_0 a$. We may then  neglect the external spring $\kappa_f$ and write the corresponding total energy as
\begin{equation}
G(\bm{x},y,t)=\sum_{i}^{N}(1+x_i)v_0+\frac{1}{2}(y-x_i)^2+\frac{\lambda_J}{2}(x_{i+1}-x_i)^2-ty.
\end{equation}
The partition function takes the form
$$
\mathcal{Z}(\beta,t) = \int dy\sum_{\{x\}}e^{-\beta G(x_i,y,t)}.
$$
The corresponding  Gibbs free energy, $\mathcal{G}(\beta,t)=-(1/(N\beta)\log \mathcal{Z}(\beta,t)$, can be again computed semi-explicitly in  the thermodynamic limit $N\rightarrow \infty$. Using the combination of a transfer matrix approach and a Laplace method,  we obtain 
\begin{equation}\label{eq:FESD}
\begin{split}
\mathcal{G}(\beta,t)=&-t y_*+v_0+\frac{y_*^2}{2}+\frac{g(y_*)}{\beta} - \frac{1}{\beta}\log\left[ \cosh g(y_*)+\sqrt{e^{-\beta \lambda_J}+\sinh^2g(y_*) } \right],
\end{split}
\end{equation}
where $g(y_*) = \frac{\beta}{2} (y_*+1/2-v_0)$, and $y_*$ is the solution of a transcendental equation 
\begin{equation}\label{eq:scsrsd}
y_*=t-\frac{1}{2}+\frac{e^{\beta  \lambda_J} \sinh g(y_*)  \sqrt{e^{-\beta  \lambda_J}+\sinh ^2 g(y_*) }}{2 e^{\beta  \lambda_J} \sinh ^2 g(y_*) +2}.
\end{equation}
\begin{figure}[ht!]
\centering
\includegraphics[scale = 1.]{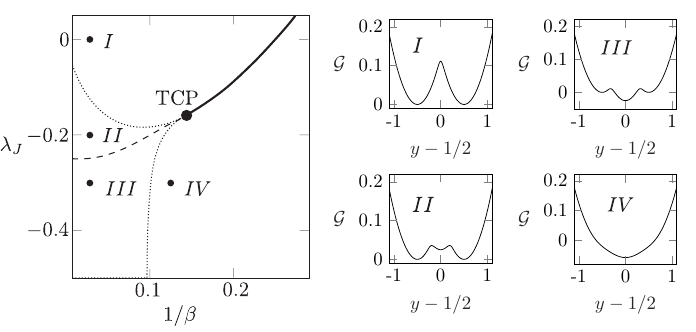}
\caption{ (a) Phase diagram for the system in the soft device at $t=0$. Parameters: $\lambda_f = 1$, $v_0=0$. The transition between the Phases I and IV is continuous (bold solid line) down to the tricritical point TCP, where it becomes first-order (dashed line). The dotted lines are spinodal boundaries. The different phases are illustrated in the right column  where we show  the dependence of the partial free energy $\mathcal{G}$ on the internal variable $y$.} 
\label{fig:PhaseDiagramSD}
\end{figure}

The parametric behavior of the marginal free energy $\mathcal{G}(\beta,y, t)$ in the stall conditions $t=0$ is illustrated  in Fig.~\ref{fig:PhaseDiagramSD}.   We again observe the same four major regimes, I, II, III and IV, which have the same meaning as in the case of hard device ensemble. However, the location of the boundary between phases and the position of the tricritical point are now different, see Fig. \ref{fig:PhaseDiagramSD}(a). In Phase III, we observe that the phase minimizing Gibbs free energy is antiferromagnetic with pre- and post-power stroke conformations of cross-bridges finely interdigitated (see below). We also see that in this phase, the macroscopically homogeneous configuration stabilizes the stall state corresponding to the minimum (rather than maximum) of the energy.

\begin{figure}[ht!]
\centering
\includegraphics[scale=1.1]{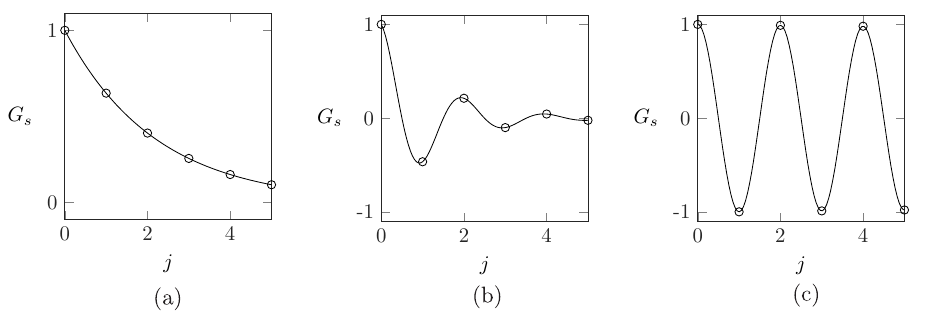}
\caption{ The equilibrium free (Gibbs) energy--strain relation in the soft device and the corresponding  tension--strain relation. (a) Phase I: $\lambda_J= 0$, $\beta = 50$; (b) Phase II: $\lambda_J = - 0.2$, $\beta = 50$; (c) Phase III: $\lambda_J = -0.4$, $\beta = 50$; and (d) Phase IV: $\lambda_J = -0.2$, $\beta = 4$.  Other parameters: $v_0 = 0$. Solid curves correspond to ground states, and dotted lines show some other stable and unstable equilibria.}
\label{fig:FE_TE_SD}
\end{figure}

As in the hard device ensemble, the parameter $\lambda_J$ plays the major role in the behavior of the system, which we illustrate by showing in Fig.~\ref{fig:FE_TE_SD} the equilibrium free energy and the force--elongation relation.  One can see that in Phases I and II, the stall state is represented by the macroscopic mixture of two pure states with all cross-bridges occupying coherently either pre- or post-power stroke configuration. Instead, in Phases III and IV, the stall state is homogeneous, and cross-bridges in different conformations are microscopically mixed. The difference again is that in Phase IV, each cross-bridge is fluctuating independently between two conformational states, while in Phase III, neighboring cross-bridges are always in different conformations that are fixed in time. 

Similarly to the case of the hard device, we can analytically find the location of the line of critical points separating Phases I and IV and determine the position of the tricritical point, see Fig.~\ref{fig:PhaseDiagramSD}(a). To define the order parameter  in the soft device case, we observe  that the equilibrium of the system with respect to the variable $y$ implies
$
y=t+ N^{-1}\sum_i x_i.
$
It is then natural  to define the order parameter again as $\phi=2\mean{x}+1$, which in the soft device case gives,
$
\phi=2 y+1-t.  
$

We again assume, for simplicity,  $v_0=0$ and focus on the stall state $t=0$.  We can then write  the partial Gibbs free energy in the form 
\begin{equation}\label{eq:FEopSD}
\mathcal{G}(\beta,\phi)=\frac{1}{8} +\frac{\phi^2}{8}-\frac{1}{\beta}\log\left[\cosh\frac{\beta\phi}{4}+\sqrt{e^{-\beta \lambda_J}+\sinh^2\frac{\beta\phi}{4}}\right].
\end{equation}
The corresponding self-consistency relation defining the equilibrium value of the  order parameter reads,
\begin{equation}\label{eq:scsropSD}
\phi_*=\frac{\sinh\frac{\beta \phi_*}{4}}{\sqrt{e^{-\beta \lambda_J}+\sinh^2\frac{\beta\phi_*}{4}}}.
\end{equation}
Near the second-order phase transition, we can expand the free energy $\mathcal{G}$ for small $\phi$ in Taylor series. Leaving only the first terms, we obtain  
$
\tilde{\mathcal{G}}(\beta,\phi)= a_0 + a_2 \phi^2+a_4 \phi^4 
$
where
\begin{subequations}\label{eq:landau_coeff_SD}
\begin{align}
&a_0 = \frac{1}{8}-\frac{\log \left(\sqrt{e^{-\beta  \lambda_J}}+1\right)}{\beta },
a_2 = \frac{1}{8}-\frac{\beta }{32 \sqrt{e^{-\beta  \lambda_J}}}, 
a_4 = \frac{\beta ^3  \left(3 e^{\beta  \lambda_J}-1\right)}{6144 \sqrt{e^{-\beta  \lambda_J}}}.
\end{align}
\end{subequations}
The critical inverse temperature, $\beta_c$, is such that the second-order term in the expansion vanishes, and hence, it must solve the equation
$
32 \sqrt{e^{-\beta_c  \lambda_J}}=8\beta_c.
$
In the limit case, where $\lambda_J\rightarrow 0$, we obtain $\beta_c=4$ which agrees with the corresponding expression in the hard device ensemble when $\lambda_f=0$.

To locate the tricritical point (TCP), we  need to extend our Landau-type expansion to the the sixth-order. We obtain
  $
\tilde{\mathcal{G}}(\beta,\phi)= a_0 + a_2 \phi^2+a_4 \phi^4+a_6 \phi^6 
$
where the new coefficient is
\begin{equation}
a_6 = \frac{ \beta ^5 \left(30 e^{\beta  \lambda_J}-45 e^{2 \beta  \lambda_J}-1\right)}{2949120 \sqrt{e^{-\beta  \lambda_J}}}.
\end{equation}

%
%
Putting to zero the second and the fourth term in this expansion, we obtain two equations  $\beta=4e^{-\beta \lambda_J/2}$  and $e^{-\beta \lambda_J/2}=\sqrt{3}$. The  tricritical point (TCP) is then located at  $\beta_{TCP}=4\sqrt{3}$ and $\lambda_{JTCP}=-\log3/4\sqrt{3}$. The first-order transition line is obtained by requiring that $\mathcal{G}(\beta,\phi=0)=\mathcal{G}(\beta,\phi=\phi_*)$, where  
\begin{equation}
\phi_* = \pm \sqrt{\frac{-a_4+\sqrt{a_4^2-3a_2a_6}}{3a_6}},
\end{equation}

\begin{figure}[ht]
\centering
\includegraphics[scale =1]{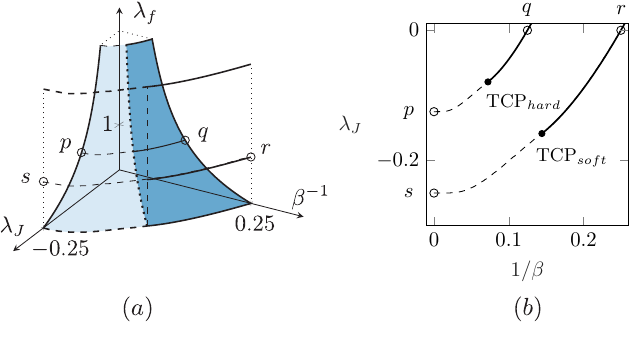}
\caption{ (a) Phase diagram  in the 3D space $(\lambda_J,  \lambda_f, \beta^{-1})$ for the case of hard device (colored surface, dark blue---second-order phase transition, light blue---first-order phase transition). The soft device phase diagram is shown by a transparent vertical cylindrical surface (b) section of both phase diagrams in (a).  For the case of a hard device, we fixed the value of parameter $\lambda_f = 1$.
}\label{fig:Double_Criticality}
\end{figure}

We can now build the full phase diagram in the 3D space  $(\lambda_J,\lambda_f, \beta)$. Phase III and IV with $\phi=0$ can be then shown separated from Phases I and II where   $\phi\neq 0$, see Fig.~\ref{fig:Double_Criticality}. We assume that the system is in the stall state which means  $z=z_0$ in the hard device and $t=0$ in the soft device ensemble.  

In the phase diagram  in Fig.~\ref{fig:Double_Criticality}(a), the plane $\lambda_f=0$ describes the  soft device ensemble, while any other plane $\lambda_f=\text{const} > 0$  corresponds to ensembles with different rigidities of the hard device. The (blue) surface separates the domain of stability of Phases III and IV  from the domain of stability of Phases I and II. In this representation, the phase boundary evolves with rigidity $\lambda_f$  producing one parametric family of ensembles. A comparison of the sections $\lambda_f=0$ (soft) and $\lambda_f=1$ (typical hard) is presented in Fig.~\ref{fig:Double_Criticality} (b)  showing, for instance,  that the location of the tricritical points in these two ensembles is different. Since for each half-sarcomere,  the value of the lump elasticity  $\lambda_f$ may vary with perpetual re-configuring of the surrounding effective matrix, the location of the actual tricritical point can be considered as floating. 

The presence, in the phase diagram of the system, of the two closely located critical lines that terminate in the two tricritical points may contribute to the system's ability to perform robustly and might be, in this sense, functional. Previously we have shown that such double criticality may actualize in the system of muscle cross-bridges due to quenched disorder \cite{BRT_PRL_2019}. Here, we neglect the quenched disorder,  which, of course, also contributes to the destabilization of the coherent response and focus instead on steric antiferromagnetic interaction as the main factor preventing strongly cooperative response. The importance of the fact that the near-criticality condition is achieved in both soft and hard device ensembles almost concomitantly stems from the mixed nature of the loading conditions experienced by a single half-sarcomere. We recall that muscle architecture involves both parallel and series connections. Parallel elements respond to a common displacement (hard device, Helmholtz ensemble), while series structures sense a common force (soft device, Gibbs ensemble). The dominance of long-range interactions  induces different collective behavior in force- and length-controlled ensembles, and to ensure the robustness of the response under a broad range of mechanical stimuli, the system can benefit from being poised in the vicinity of both types of critical regimes.

Note also that  throughout this paper, we have been assuming that the stiffness $\lambda_f = \kappa_f /N$ is a size-independent constant. Therefore, we implicitly assumed that $\kappa_f  \sim N$. We recall that the  (macroscopic) stiffness   $\kappa_f$ represents  elastic filaments which support a  parallel bundle of $N$ cross-bridges. An alternative assumption would be that filaments are not stiffer than cross-bridges and therefore $\kappa_f$ is $N$ independent, see, for instance, \cite{BRT_PRL_2019}. In this case, $\lambda_f \sim 1/N$, and the number of cross-bridges $N$ becomes a control parameter in the  phase diagram. This  means that an active control of the number of attached myosin heads could induce a structural change in the overall mechanical response of the system. For instance, it was shown that the fraction of myosin motors attached in the OFF state (heads folded back toward the M-line) depends on the thick filament stress \cite{Linari_Nature_2015}. The time course of the change in conformation of the population of myosin motors in a half-sarcomere was also traced by quick freezing of a single muscle fiber and by using cryo-EM \cite{Taylor_Cell_1999}. Such modulation of the number of attached cross-bridges  may be an indication of an  active control; however, the theoretical exploration of this idea will not be  pursued in this paper.

\section{Antiferromagnetic order}

Note that the order parameter $\phi$, which is the mechanical analog of magnetization, cannot differentiate between Phases I and II when $\phi \neq 0$ and III and IV when $\phi = 0$. Finer features of the local spin distribution can be captured if we introduce parameter distinguishing between ferromagnetic and antiferromagnetic spins' arrangements.

We recall that in magnetism, the system is antiferromagnetic if it is energetically favorable for neighboring spins to align in opposite directions. In the simplest antiferromagnets, the crystal may be divided into two sublattices, A and B, so that if the spins occupying one sublattice point one way, those occupying the other point the opposite way so that the spins of nearest-neighbor atoms are always antiparallel \cite{crangle2012solid, Zvyagin_LTP_2006}. Since the net magnetization is equal to zero in such a state, the order parameter $\phi$ cannot distinguish between paramagnetic (our Phase IV)  and antiferromagnetic (our Phase III) phases.

To obtain an approximate picture of the transition to antiferromagnetic behavior, we employ a two-sublattice mean-field approximation \cite{Nagle_JAP_1971, Agra_2006, Nishino_PRB_2016}. We first introduce the conventional spin variable $s_i = 2 x_i +1$, so that $s_i = \pm 1$, and then distinguish magnetizations in different sublattices introducing  $\phi_a = \mean{s_{i\in A}} =  (2/N) \sum_{i\in A} s_i,$  and $\phi_b =\mean{sz_{i\in B}}= (2/N) \sum_{i\in B} s_i.$  Then, the \emph{staggered} magnetization can be defined as $\phi_{s} = (\phi_a - \phi_b)/2,$ while $\phi  = (\phi_a + \phi_b)/2$.

\begin{figure}[ht]
\centering
\includegraphics[scale = 1]{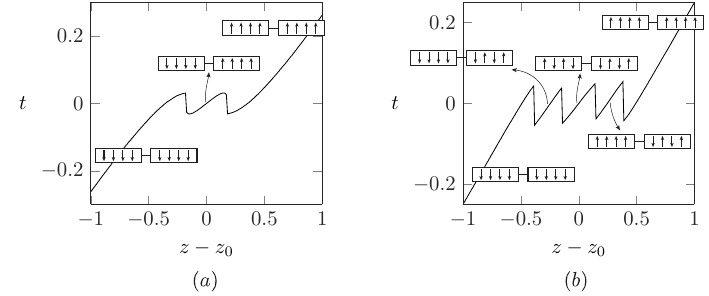}
\caption{(a) Phase diagram of a model mean-field system in a hard device showing the boundary separating  domains with different patterning of  spins (representing  pre- and post-power stroke configurations of the myosin heads). In the paramagnetic (PM) domain,  spins are disordered; in the antiferromagnetic (AFM) domain, spins are ordered (interdigitated).  The loading parameter is fixed at  $z= z_0^s$. (b) Staggered magnetization $\phi_s$ as a function of the non-dimensional temperature $1/\beta$.  (c) Order parameters for the AFM phase $\phi_{a,b}$ as  functions of   $1/\beta$.  Other parameters:   $\lambda_J = -0.5$. }
\label{fig:AFM_PhaseDiagram}
\end{figure}

If we express  the Hamiltonian \eqref{eq:Hamiltonian} in terms of the spin variables, $ s_i=\pm 1$, the relaxed energy can be written as,
\begin{equation} 
\begin{split}
E(s_i,z)=-\frac{1}{8 N (1+\lambda_f)}\sum_{i,j} s_i s_j - \frac{\lambda_J}{4}\sum_i s_{i}s_{i+1}-h(z) \sum_i  s_i +f(z).
\end{split}
\end{equation}
where $$h(z)=\frac{2\lambda_f z-1}{4(1+\lambda_f)}+\frac{1}{4}-\frac{v_0}{2}+\frac{\lambda_J}{2},\,\, f(z)=\sum_i \frac{\lambda_f z(1+z)}{2(1+\lambda_f)}+\frac{1}{4}+\frac{v_0}{2}-\frac{1}{8(1+\lambda_f)}+\frac{\lambda_J}{4}.$$
This  energy can be  subdivided into two lattices (A and B), and in  the mean-field approximation,  we can  express the values of spins on the corresponding sublattices as $ \phi_{a,b} +  \tilde{s}_{i\in A,B}$, where $\phi_{a,b}  $ are the average values   and  the fluctuations $  \tilde{s}_{i\in A,B} $  are considered to be small. Then, we obtain the    approximate Hamiltonian  
\begin{equation}
\begin{split}
\mathcal{H}_{mf} & = -\frac{1}{8 N (1+\lambda_f)} \sum_{i,j} (\phi_a +\tilde{s}_{i\in A} +\phi_b +\tilde{s}_{i\in B} )(\phi_a +\tilde{s}_{j\in A} +\phi_b +\tilde{s}_{j\in B} )
\\
&-\frac{\lambda_J}{4}\sum_{<i,j>} (\phi_a +\tilde{s}_{i\in A} +\phi_b +\tilde{s}_{i\in B} )(\phi_a +\tilde{s}_{j\in A} +\phi_b +\tilde{s}_{j\in B} )
\\
&-h(z) \sum_{i\in A} (\phi_a +\tilde{s}_{i\in A}) -h(z) \sum_{i\in B} (\phi_b +\tilde{s}_{i\in B}) +f(z),
\end{split}
\end{equation}
where the notation $<i,j>$ indicates that $i$ and $j$ are nearest neighbors. We neglect quadratic terms in the fluctuation, and then, the problem is posed  for $s_{i\in A,B}$, by substituting $\tilde{s}_{i\in A,B} = s_{i\in A.B}-\phi_{a,b}$,
\begin{equation}
\mathcal{H}_{mf} = N J_{eff} \phi_a\phi_b + h_a(z) \sum_{i \in A} s_i +h_b(z)\sum_{i\in B}s_i +c(z),
\end{equation}
where $$J_{eff} = \frac{1}{4} \left(\frac{1}{\lambda_f+1}+2 \lambda_J \right)$$ is the effective (mean-field) coupling, which depends on  both short- and long-range interactions. The loading-dependent effective fields  are 
 $h_a(z) = -\phi_b J_{eff} - h_s(z)$ and $h_{b}(z) = -\phi_a J_{eff} - h_s(z) $, with  
 $$h_s(z)=\frac{2\lambda_f z-1}{4(1+\lambda_f)}+\frac{1}{4}-\frac{v_0}{2}+\frac{\lambda_J}{2}.$$ 
Finally, the additive constant is   $c(z) =  \frac{N\lambda_fz^2}{2(1+\lambda_f)} +N v_0$. 
Using the self-consistency conditions   $\phi_{a,b}=<\sum_{i\in {A,B}} s_i>$, we obtain  the  nonlinear algebraic equations for $\phi_{a,b}$:
%
%
\begin{equation}\label{eq:AFM_magnetization}
\begin{split}
\phi_a = \frac{1}{\mathcal{Z}_{mf}}\sum_{\{s_i\}} s_{i\in A} e^{-\beta \mathcal{H}_mf} =  \tanh\beta (J_{eff} \phi_b  + h_s) \\
\phi_b = \frac{1}{\mathcal{Z}_{mf}}\sum_{\{s_i\}} s_{i\in B} e^{-\beta \mathcal{H}_mf} = \tanh\beta (J_{eff} \phi_a  + h_s)
\end{split}
\end{equation}
where $
\mathcal{Z}_{mf} = 2^N e^{-N\beta J_{eff}\phi_a\phi_b}\left( \cosh \beta h_a \cosh \beta h_b\right)^{N/2}.
$ The system \eqref{eq:AFM_magnetization}
  has a unique solution $\phi_a = \phi_b = 0$ for  $\beta \leq \beta_N = - 1/J_{eff}$ and acquires two additional  nonzero solutions for $\beta > \beta_N$.  The analog of stall state  in this mean-field problem is  $z=z_0^s=(1+1/\lambda_f) (v_0-\lambda_J)-1/2$. The phase diagram at $z=z_0^s$  is presented in Fig.~\ref{fig:AFM_PhaseDiagram} where we set for simplicity that  $v_0 = 0$ and $\lambda_f = 1$. 

Since we neglected the crucial long-range interactions in our system due to myosin backbones and treated the steric short-range interactions only at the mean-field level, the resulting phase diagram in Fig.~\ref{fig:AFM_PhaseDiagram}(a) is too oversimplified to capture our four  Phases  I, II, III and IV.  The only division which can be illustrated in this way is between the paramagnetic (PM) Phase IV and antiferromagnetic (AFM) Phase III. The corresponding line on the plane $(\lambda_J, 1/\beta)$   clearly shows that the AFM phase is favored by low temperatures and strong negativity of the coupling constant $ \lambda_J$. 
 
To reveal the antiferromagnetic ordering emerging in Phases II and  III, we can also characterize the two-point correlation function directly 
$G(i,i+j) = \mean{x_i x_j}  - \mean{x_i}\mean{x_j}$  in the hard device ensemble by using again the transfer matrix method. 
%
Following \cite{goldenfeld1992lectures}, we write 
\begin{equation}
\begin{split}
\mean{x_i} =& \frac{1}{\mathcal{Z}} \int \, dy \sum_{\{x\}}  e^{-\beta E(x_i,y,z)} \,x_i
 = \frac{1}{\mathcal{Z}} \int \, dy e^{-N\beta \varphi (y,z)}  
 \sum_{x_i} \cdots \sum_{x_N}\left[ \bm{T}_{x_1 x_2} \bm{T}_{x_2 x_3}  \cdots \boldsymbol{T}_{x_{i-1} x_i} x_i \,\boldsymbol{T}_{x_{i} x_{i+1}} \cdots \right]
\end{split}
\end{equation}
Note that the matrix $\sum_{x_i}\boldsymbol{T}_{x_{i-1} x_i} x_i \,\boldsymbol{T}_{x_{i} x_{i+1}} $  can be  equivalently written as 
$
\boldsymbol{A} = \boldsymbol{T}\boldsymbol{X}  \boldsymbol{T}
$,
where $\boldsymbol{X} = \left(\begin{array}{cc}
0 & 0\\ 
0 & -1
\end{array} \right)$ and therefore 
$
\mean{x_i} = \frac{1}{\mathcal{Z}} \int \, dy \, e^{-N\beta \varphi (y,z)}\Tr (\boldsymbol{X} \boldsymbol{T}^N).
$
However,  $\boldsymbol{T}= \boldsymbol{S} \boldsymbol{T ' }\boldsymbol{S} ^{-1}$, where $\boldsymbol{T '}$ is a diagonal matrix with eigenvalues, given by Eq.~\eqref{eq:eigenvalues}. Then, 
$
\mean{x_i} = \frac{1}{\mathcal{Z}} \int \, dy\,  e^{-N\beta \varphi (y,z)}\Tr [\boldsymbol{S}^{-1}\boldsymbol{X} \, \boldsymbol{S}\,(\boldsymbol{T '})^N].
$ We can explicitly compute the eigenvectors  of $\boldsymbol{T}$, $V_{\pm} = (\alpha_{\pm}, 1)$, where $\alpha_\pm = e^{\frac{\beta J}{2}}[-\sinh h\pm \sqrt{e^{-\beta J}+\sinh ^2 h}]$, so  $\boldsymbol{S} =\left(\begin{array}{cc}
\alpha_+ & \alpha_-\\ 
1 & 1
\end{array} \right)$, and 
\begin{equation}
\boldsymbol{S}^{-1}\boldsymbol{X} \, \boldsymbol{S} =
\left(\begin{array}{cc}
\chi_{11}  & \chi_{12}\\ 
\chi_{22} & \chi_{22}
\end{array} \right)
\equiv \left(\begin{array}{cc}
 -\frac{\sinh h}{2\sqrt{\sinh ^2 h+e^{-\beta J}}}- \frac{1}{2} \quad 
 & \frac{e^{-\frac{\beta  J}{2} }}{2 \sqrt{\sinh ^2 h+e^{- \beta  J}}}\\ 
\frac{e^{-\frac{\beta  J}{2} }}{2 \sqrt{\sinh ^2 h+e^{-\beta J}}}\quad & \frac{\sinh h}{2\sqrt{\sinh ^2 h+e^{-\beta  J}}}- \frac{1}{2}
\end{array} \right).
\end{equation}
Then, using Laplace method and the fact that $\lim_{N\to \infty}(\lambda_2/\lambda_1)^N \to 0$, we obtain 
\begin{equation}
\mean{x_i} = \chi_{11}=  -\frac{\sinh h}{2\sqrt{\sinh ^2 h+e^{-\beta J}}}- \frac{1}{2}.
\end{equation}

\begin{figure}[ht!]
\centering
\includegraphics[scale = .95]{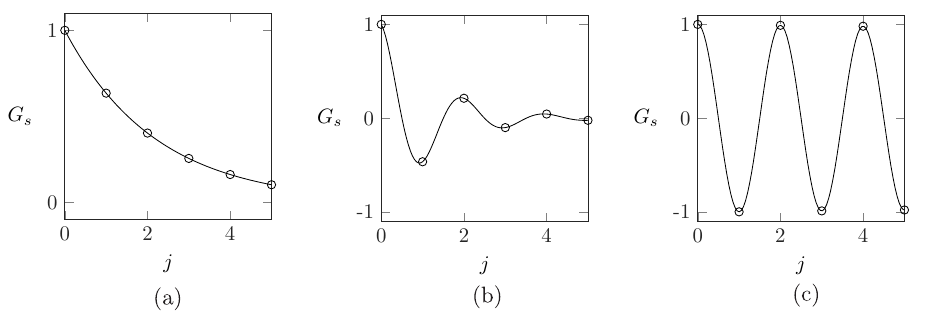}
\caption{ Two-point correlation function in different phases  for the model system, see  Fig.~\ref{fig:AFM_PhaseDiagram}(a).  The loading parameter is fixed at  $z= z_0$.   Other parameters: (a) $\beta = 20$, $\lambda_J = 0.1$ PM-phase; (b) $\beta = 10$, $\lambda_J = -0.2$ PM-phase; and (c) $\beta = 20$, $\lambda_J = -0.5$ AFM-phase.} \label{fig:Correlations}
\end{figure}

The two-point correlation function can be compute similarly. Note first that   
\begin{equation}
\begin{split}
\mean{x_i x_j} = \frac{1}{\mathcal{Z}} &\int \, dy \, e^{-N\beta \varphi (y,z)}   
 \Tr\left[ \boldsymbol({S}^{-1}\boldsymbol{X} \, \boldsymbol{S})\,(\boldsymbol{T '})^j (\boldsymbol{S}^{-1}\boldsymbol{X} \, \boldsymbol{S})\,(\boldsymbol{T '})^{N-j}\right]
\end{split}
\end{equation}
Again, in the limit $N\to \infty$, we obtain 
$
\mean{x_i x_j} = \chi_{11}^2+\chi_{12}\chi_{21}\left(\lambda_2/\lambda_1\right)^j,
$
which allows us to finally write 
\begin{equation}
\label{G} 
G(i,i+j)  = \mean{x_i x_j}  - \mean{x_i}\mean{x_j}
  = \chi_{12}\chi_{21}\left(\frac{\lambda_2}{\lambda_1}\right)^j = \frac{1 }{4+4\, e^{ \beta  J} \sinh ^2 h}\left[\frac{\cosh h-\sqrt{e^{-\beta  J}+\sinh ^2 h}}{\cosh h+\sqrt{e^{-\beta  J}+\sinh ^2 h }} \right]^j.
\end{equation}
Here, $h = \frac{\beta}{2}(y_*+1/2-v_0)$, and $y_*$ is given by \eqref{eq:scsr}. To illustrate \eqref{G}, 
%
%
%
it is convenient to use the spin variables  $s_i = \pm 1$  defined by $2x_i=s_i-1$. In Fig.~\ref{fig:Correlations}, we  show  the function $G_s(j) \equiv G_{s_i} (i,i+j) = 4 \,G_{x_i}(i,i+j)$ which clearly distinguished between the ferromagnetic (FM) and antiferromagnetic (AFM) phases; note that we do not attempt here to directly associate these regimes  with the phases defined in Fig.~\ref{fig:PhaseDiagram}.

%
%

To summarize, we have shown if short-range steric interactions between cross-bridges of antiferromagnetic type are sufficiently strong, then the state of isometric contractions is not only mechanically stable in the sense that it has positive stiffness, but it is also ordered in the way that exactly half of the cross-bridges is in pre-power stroke state and another half in the post-power stroke state; moreover, the conformation states oscillate at lattice scale so that both confirmations are effectively present at any location inside a half-sarcomere.

\section{Two half-sarcomeres}

In this section, we show how the presence of antiferromagnetic short-range interactions allows the system to avoid instability developing into macroscopic inhomogeneity. One can say that to maintain homogeneity at the macroscale, such system chooses to be maximally inhomogeneous at the microscopic scale.

So far, we have been dealing with a single half-sarcomere behavior, which is the basic unit of contraction. Even having non-convex elastic energy and negative stiffness, such a unit is still stable in a hard device. This case's instability comes when more than one element with non-convex energies is loaded in a hard device while being arranged in series. Even if the system includes only two units like this, they do not need to deform in an affine way and can accept instead different deformational states.

\begin{figure}[ht]
\centering
\includegraphics[scale =1]{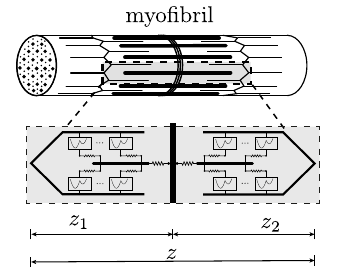}
\caption{Schematic representation of a single sarcomere. Inset illustrates   the  equivalent mechanical model:     two half-sarcomeres connected in series. }\label{fig:SeriesStructure}
\end{figure}

This is an important issue since muscle fiber is often represented as a  series of connections of half-sarcomeres. It has been shown that if the energy of individual half-sarcomeres is non-convex, such association breaks into macroscopic islands of either fully pre- or fully post-power stroked half-sarcomeres \cite{Puglisi2000, Vilfan_BioPhys_2003}. Such instability would be prevented if the energy of a half-sarcomere is convexified in the stall configuration. As we have shown in the previous sections, such convexification occurs when the system is either in Phase IV or Phase III. It has been already known to Huxley that the first option is not viable in view of insufficiently high reduced temperature in physiological conditions.  That is why we consider below the only low-temperature option when the destabilizing steric short-range interactions are sufficiently strong, and the physiological system is in Phase III.

To illustrate  the disappearance of the macroscopic  instability in this case, it is sufficient to consider the simplest system  with  only two half-sarcomeres connected in series. Such system can be viewed as a prototypical description of a single sarcomere, see Fig.~\ref{fig:SeriesStructure}.  Experiments suggest that in stall conditions, each  sarcomere  contains a macroscopically (but not necessarily microscopically) uniform mixture of pre- and post-power stroke cross-bridges \cite{MIDDE20111024}.   

Suppose that each half-sarcomere is equilibrated independently at the microscale and therefore behaves at the macroscale according to the elastic constitutive relation generated by the free energy $\mathcal{F}(\beta,z)$, see \eqref{eq:FE}.  We can then write the equilibrium energy of the whole system loaded in a hard device  in the form
\begin{equation}
\label{E_2}
 \mathcal{F}_2(\beta, z, z_1) = \mathcal{F}(\beta,z_1)+\mathcal{F}(\beta, z-z_1).
\end{equation}
where $z$ is the macroscopic controlling parameter. The extra variable $z_1$ can be eliminated using the equilibrium condition $\partial \mathcal{F}(\beta,z_1) / \partial z_1 =  \partial \mathcal{F}(\beta,z-z_1) / \partial z_1$.  Substituting the   equilibrium value 
$z_1(\beta, z)$ into \eqref{E_2}, we  obtain the function $ \mathcal{F}_2(\beta, z) = \mathcal{F}_2(\beta, z, z_1(\beta, z,))$ and construct the equilibrium force--elongation relation $t(\beta, z) =d \mathcal{F}_2(\beta, z) /dz$. It  would be  also of interest to compute for each half-sarcomere, the fraction of  cross-bridges  in either pre- or post-power stroke conformations. For half-sarcomeres 1 and 2, respectively, it will be given by the functions 
\begin{equation}
\phi_{1,2} (\beta, z) = 2(\lambda_f+1)y_*^{1,2}(\beta, z) - 2\lambda_f z+1
\end{equation}
where the function  $y_*(\beta, z)$ is given by  \eqref{eq:scsr}.
\begin{figure}[ht]
 \centering
\includegraphics[scale =1]{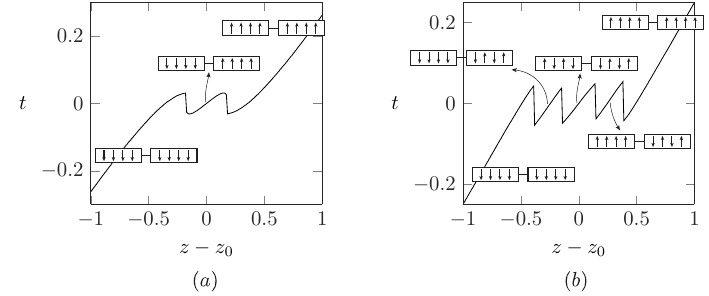}
\caption{Mechanical response of the system consisting of two half-sarcomere units connected in series with each half-sarcomere represented by infinite number of cross-bridges (thermodynamic limit). Solid curves show the equilibrium tension--elongation curves; the insets illustrate the associated  configurations of cross-bridges. Parameters: (a)  $v_0=0$, $\beta = 5$, $\lambda_f = 1$ and $\lambda_J = 0$ (no short-range interactions). (b) $v_0=0$, $\beta = 25$, $\lambda_f = 1$ and $\lambda_J = -0.25$ (strong antiferromagnetic short-range interactions).}\label{fig:SeriesTension}
\end{figure}

\begin{figure}[ht]
 \centering
\includegraphics[scale =1]{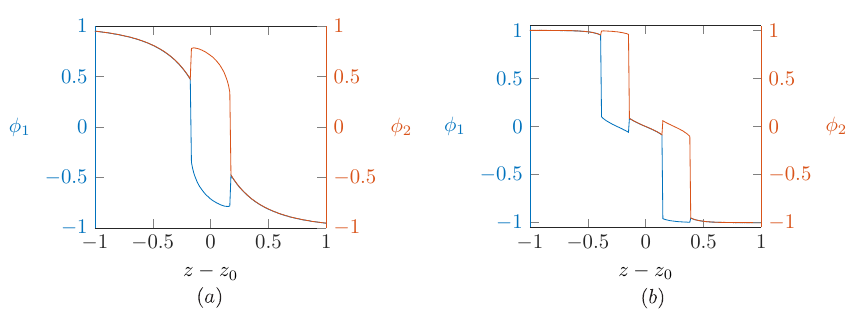}
\caption{Strain response of the 'local' order parameters for the system of two half-sarcomeres connected  in series. The colored curves show $\phi_1$ and $\phi_2$ are the order parameter of the individual half-sarcomeres. Parameters:  (a): $v_0=0$, $\beta = 5$, $\lambda_f = 1$ and $\lambda_J = 0$ (no short-range interactions); b)  $v_0=0$, $\beta = 25$, $\lambda_f = 1$ and $\lambda_J = -0.25$ (strong antiferromagnetic  short-range interactions).}\label{fig:SeriesPhi}
\end{figure}

In Fig.~\ref{fig:SeriesTension} and Fig.~\ref{fig:SeriesPhi}, we compare  the response of two half-sarcomeres  in series depending on whether their constitutive response follows the one from  Phase I, see  Fig.~\ref{fig:SeriesTension}(a) or Phase III, see Fig.~\ref{fig:SeriesTension}(b). 

If  short-range interactions are of \emph{ferromagnetic } type, and the individual subsystems are in Phase I, we observe that while the effective stiffness in the stall state $z=z_0$ is positive, see Fig.~\ref{fig:SeriesTension}(a), the macro-state is non-affine  with  $\phi_1$ and $\phi_2$ taking different values. This means that the two half-sarcomeres in series  have different internal structures that are both pure: One is fully in pre-power stroke state, and another one is fully in post-power stroke state, see Fig.~\ref{fig:SeriesPhi}(a). 

If instead, the short-range interactions are of \emph{antiferromagnetic} type  and the individual half-sarcomeres are in Phase III, the whole sarcomere (in stall conditions)  has   positive stiffness, see  Fig.~\ref{fig:SeriesTension}(b), and its macro-configuration is affine  with  the order parameters $\phi_1$ and $\phi_2$ taking the same value, see  Fig.~\ref{fig:SeriesPhi}(b). This means that the two half-sarcomeres in series have the same internal structure, which is, however, micro-inhomogeneous, with cross-bridges  equally divided between pre- and post-power stroke conformations.  
 
Outside a neighborhood of the stall state, the system with \emph{ferromagnetic}-type short-range interactions is microscopically homogeneous with all half-sarcomeres occupying the same pure state. Instead, the system with \emph{antiferromagnetic}-type interactions develops two zones  of non-affine behavior where one of the two half-sarcomeres is in a pure state, while the other one is in a mixed state.


This simplified description can be immediately extended to cover the case of $N$ half-sarcomeres connected in series. In the case of dominating antiferromagnetic interactions, the stall state will remain mechanically stable, macroscopically homogeneous and microscopically inhomogeneous. Outside the stall state's immediate vicinity, there will be different regimes with a variable number of coexisting half-sarcomeres in a pure and mixed state. The overall macroscopic force--elongation curve will then be wavy with an overall slightly positive average slope as observed in experiments.  In the limit $N \to \infty$, we obtain weak convergence to an apparently stress--strain curve whose local microscopic derivative is nevertheless different from its macroscopic (averaged) value.

\section{Conclusions} 
 
We presented a mathematical model describing the passive mechanical response of tetanized muscle fibers. It is relevant for analyzing experiments involving fast mechanical perturbations imposed on the state of isometric contractions.  The observed overall homogeneity of the \emph{macroscopic}  response in these conditions was previously shown to contradict the \emph{microscopic} prediction of  the classical HS model and its more recent generalizations \cite{CT_ROPP_2018}. According to this model the slope of the force--length curve should be negative  and the  cross-bridges inside single half-sarcomeres must be necessarily either all in pre- or all in post-power stroke states.   

The standard HS-type model assumes that the bi-stable nature of myosin heads can be modeled by spin variables and that inside a single half-sarcomere, the myosin cross-bridges are arranged in parallel  \cite{Irving02, Higushi_Nature2017}. This parallel structure is responsible for mean-field-type interaction among different cross-bridges, which forces them to act synchronously. To compromise the coherent response of such parallel bundles, we proposed to move beyond the HS framework  \cite{CAT_PRL} and introduce into the model the antiferromagnetic mechanical interactions between neighboring cross-bridges. We then studied the competition between these destabilizing short-range interactions and the stabilizing long-range interactions of HS.

We showed that while an increase in the strength of the ferromagnetic short-range interactions ($\lambda_J>0$) would have the same effect as a decrease in temperature, leaving the overall qualitative behavior of HS unchanged,   the presence of antiferromagnetic interactions  ($\lambda_J<0$) drastically changes the qualitative behavior of the system. The main effect is that by tuning the parameter $\lambda_J<0$, one can introduce a new energy well corresponding to the stall state and, in this way, stabilize the state of isometric contractions. At a fixed strength of long-range interactions (fixed $\lambda_f$), the phase diagram in $(\beta,\lambda_J)$ space shows a  line of second-order phase transition which expands the critical regime found in \cite{CAT_PRL} at $\lambda_J=0$.  

The proposed model shows that to make a macroscopically homogeneous response possible, the system must be microscopically inhomogeneous. Micro-inhomogeneity, achieved by activating short-range interactions of antiferromagnetic type, takes the form of periodic interdigitated patterns with neighboring sarcomeres taking alternatively either pre- or post-power stroke conformations. Such patterns replace the highly coherent micro-homogeneous mechanical response favored by the original HS model. The predicted micro-inhomogeneity may be beneficial as it breaks the coherency of the response and facilitates the swift transition to either fully pre- or fully post-power stroke configuration in response to external excitation.  We also argue that evolution could use short-range interactions to tune the muscle machinery to perform near the conditions of 'double criticality', where both the Helmholtz and the Gibbs free energies are singular. Such design would be highly functional given that elementary force-producing units usually perform in a mixed soft--hard loading conditions.  

We have shown that the proposed hypothesis of the 'interacting neighbors'  questions the stability of the uniform equilibrium (micro)configurations of half-sarcomeres only in the case when they respond to the mechanical perturbation analyzed in the classical HS paper. More theoretical studies are needed to evaluate how it would also affect the \emph{general} kinetic response of muscle myofibrils in both passive and active phases.   Another challenge is to assess how the micro-inhomogeneous nature of the response would influence the ability of the motors to use the chemical energy of the ATP. In the HS type models, this energy can be linked to the value of $v_0$, which may be then affected by the presence of short-range interactions.  Yet another important factor to be included in future studies is the complex 3D arrangements of myosin fibers and the presence of an elaborate 3D architecture linking individual cross-bridges not only through thick and thin filaments but also, effectively, through  M- and Z-disks.  Only by surviving such more realistic tests in fully comprehensive settings can the hypothesis of antiferromagnetic short-range interaction gain acceptance in the muscle mechanics community.

\section{Acknowledgments } 
The authors thank Marco Linari and Matthieu Caruel for helpful discussion. H.B.R. was supported by a PhD fellowship from Ecole Polytechnique; L. T. was supported by the grant  ANR-10-IDEX-0001-02 PSL.

\addcontentsline{toc}{section}{References}

\end{document}